%% file: surf3.tex
\documentclass[12pt,a4paper]{article}
\usepackage[]{epsfig}	      
\usepackage[T1]{fontenc}      
\usepackage[latin1]{inputenc} 
\usepackage[english]{babel}    
\usepackage[]{latexsym} 

\def\zfo{${{\mbox{$\bigcirc$}}\!\!\!\!\!\!\:{\mbox{2}}\,}$} 
\def\ffo{${{\mbox{$\bigcirc$}}\!\!\!\!\!\!\:{\mbox{5}}\,}$} 
      
\begin{document}
\thispagestyle{empty}
\title{Tiling theory applied to the surface structure of 
icosahedral $AlPdMn$ quasicrystals.}
\author{ Peter Kramer, Zorka Papadopolos, Harald Teuscher\\
Institut für Theoretische Physik der Universit\"at
T\"ubingen}

\maketitle
\section*{Abstract.}

Surfaces in i-$Al_{68}Pd_{23}Mn_9$ as observed with STM and LEED experiments
show atomic terraces in a Fibonacci spacing. We analyze them in a bulk
tiling model due to Elser which incorporates many experimental data. 
The model has 
dodecahedral Bergman clusters within an  icosahedral 
tiling ${\cal T}^{*(2F)}$ and is projected from
the $6D$ face-centered hypercubic lattice.
We derive the occurrence and Fibonacci spacing of atomic planes 
perpendicular to any 5fold axis, compute the variation of  
planar atomic densities, and determine the (auto-) correlation functions.
Upon interpreting the planes as terraces at the surface we find 
quantitative agreement with the STM experiments.

\section{Introduction.}

The bulk structure of the icosahedral phases i-$AlPdMn$, i-$AlFeCu$ 
and their modelling  in terms of a $6D$ description 
has been an active research field for more than one decade. 
From the
many papers on this topic we mention only a few. Many more 
references are quoted in these publications. 
Katz and Gratias
\cite{KA} derive from previous work for i-$AlFeCu$ a quasiperiodic 
network of atomic positions. It is  generated by three basic atomic 
windows related to the $6D$ hypercubic F-lattice.
They examine carefully the interatomic distances. De Boissieu et al. 
\cite{BO} determine for i-$AlPdMn$ from X-ray and neutron data 
in detail the decomposition of
the atomic surfaces. All these models use the $6D$ embedding,
the parallel and the perpendicular space.
Elser \cite{elser} 
generalizes  and unifies both 
these models and interprets them in terms of clusters occupying
the odd and even vertices of the icosahedral tiling related to the 
$6D$ hypercubic P-lattice: The odd vertices  
form the centers of Bergman clusters, which then around the even
vertices build up Mackay clusters. Additional 
atomic positions are related to this basic structure. The Elser 
model was actually created for the study of random tilings, but
by construction admits a perfect tiling structure which then 
incorporates the main experimental 
data which have led to the models by Katz and Gratias 
and by de Boissieu et al.
As shown in 
\cite{KR1}, the Elser  model
can be taken as a {\em network of atomic positions in  a tiling model}, 
denoted by
${\cal T}^{*(2F)}$ and related to the observed hypercubic F-lattice
and -module. This tiling model will be used in what follows.
Its composite atomic surfaces are 
closely related to those of the Katz-Gratias model \cite{KA}.

\vspace{0.2cm}

The surface structure of i-$Al_{68}Pd_{23}Mn_9$ perpendicular 
to 5fold axes has been explored by various groups. Schaub et al.
\cite{BU} applied scanning tunneling microscopy (STM) and
low-energy electron diffraction (LEED) to obtain atomic scale
information of a sputtered and annealed surface. They observe
a sequence of 11 atomically flat terraces. Two spacings of 
(4.22 $\pm$ 0.26) and (6.78 $\pm$ 0.24) \AA $\;$  form a 
Fibonacci string of the type LLSLLSLSLL.
Pentagonal holes of a single fixed orientation appear 
within these terraces. 
Gierer et al \cite{GI1,GI2} by dynamical LEED studies 
for a similarly prepared 
surface confirm the quasicrystalline structure. To interpret  their data 
they perform  dynamical diffraction calculations for assumed terminations
of a model patch from the bulk model of de Boissieu et al. \cite{BO}. They find 
optimal agreement for Al-rich terminations of high atomic density.
A study by Ebert et al. \cite{EB} 
of in-situ cleaved surfaces revealed terraces only after annealing
of the initially rather rough surface. 
\vspace{0.2cm}

For the theory of quasicrystals, the experiments raise the question what 
quasiperiodic repetition
pattern and what structure variation within planes
can be rigorously obtained from a bulk model  of i-$Al_{68}Pd_{23}Mn_9$. 
In the present paper we address these questions
in the  description by the tiling model.
We develop a quasiperiodic analysis similar to the one of crystal surfaces 
in terms of particular netplanes.  To obtain exact results we cannot 
rely on features seen in a model patch. 
Instead we make full use of the 
technique of windows or coding for quasiperiodic structures. The general 
principles of the window technique have been described in several
monographs on quasicrystals. We illustrate them on the 
well-known one-dimensional Fibonacci paradigm.  
We then apply the unique lifting and projection 
method between the physical (parallel)  and the window (perpendicular)
space, called the star-map by Moody \cite{MO}, 
to the icosahedral F-module, to
the tiling,  and to its decoration.
It turns out that our main results can be expressed 
in terms of the one-dimensional Fibonacci system.
\vspace{0.2cm}

We now survey  the model input  and the content of the following sections.
Our bare  tiling model has the following data:
We adopt the ${\cal T}^{*(2F)}$ tiling model
projected from the face-centered hypercubic lattice (2F) $\sim D_6$ in $E^6$. 
Upon scaling by a  factor $2$, the lattice (2F)  
comprises the {\em even vertex points (even index sum)} 
of the full hypercubic lattice $P$ whose projection was given in \cite{KR0}.  
For a full description of the tiling and its projection we refer to \cite{KR}. 
Its vertex points are projected lattice points.
We shall use two units of length: 
\ffo $\;$ is the length along 5fold axes of the six basis 
vectors $e_i, \, i=1,\ldots , 6$ of the hypercubic lattice, 
projected to the two invariant icosahedral
subspaces $E_{\parallel}$ or $E_{\perp}$ respectively. Along projected 
2fold axes we choose
the standard length \zfo$\; = \frac{2}{\sqrt{\tau+2}}$\ffo. To convert 
to atomic distances in i-AlPdMn  we adopt from \cite{elser, KR1}
the $\tau$-scaled short edge length of the tiling, 
\begin{equation}
s = \tau \mbox{\zfo} = \frac{2\tau}{\sqrt{\tau+2}} \mbox{\ffo},
\;  \mbox{\ffo} \rightarrow 4.56 \; \mbox{\AA}. 
\end{equation}
The window of the vertex points for the tiling is in $E_{\perp}$ the 
triacontahedron \cite{KE}, \cite{KR} shown in Fig.5. 
The tiling is decorated according to Elser \cite{elser} 
with dodecahedral Bergman clusters \cite{KR1}. The midpoints of these
Bergman clusters are placed on the projected {\em odd vertex points}
of the hypercubic lattice. Their edge length is  
$\tau^{-1}$ \zfo $\; = 2.96$ \AA $\;$, their
height along a 5fold direction is $\frac{2\tau^2}{\tau+2}$ \ffo $\; = 6.60$ \AA $\;$. 
For all other atomic positions, most of which do 
not enter the present analysis, we refer
to \cite{elser, KR1}. 
\vspace{0.2cm}

In section 2 we develop the window technique for the bulk tiling
and its planes perpendicular to 5fold axes. 
We start in 2.1 with the Fibonacci tiling  and explain 
the technique of windows. We briefly describe the icosahedral tiling
${\cal T}^{*(2F)}$ for the F-phase, section 2.2, and planes of vertex points 
perpendicular to a 5fold axis in a 3D space $E_{\parallel}$, section 2.3,
and give their windows in $E_{\perp}$. We shift between 
these planes along Fibonacci lines, section 2.4, and show that
most of the vertex points belong to a system of shifted planes, section
2.5. In section 2.6 we interpret the terraces found in the STM
experiment as terminations of the bulk model. From the tiling model 
we prove the existence of a full Fibonacci sequence of planes and
of a spacing as found in the STM experiment, and we
predict variations of the density of vertex points along the sequence,
with bounds from the observed Fibonacci string. 
\vspace{0.2cm} 

In section 3 we use the decoration of the tiling to infer 
more structure information within the planes from other atomic positions
of the tiling model. In particular we look for
pentagonal structures as seen in the STM experiments \cite{BU}.
We consider the dodecahedral Bergman clusters of the Elser model \cite{elser} 
on the tiling. The dodecahedra have two pentagonal vertex sets 
of the same orientation perpendicular to a 5fold axis. The corresponding
cutting planes are transformed in subsection 3.1 by lifting and
projection to the perpendicular 
space. Their window description with respect to the triacontahedron
is derived. The correlation with vertex points of the tiling gives rise
to three alternative models for the structure within planes. 
The predicted density of
vertex points and pentagons is derived in exact form in section 3.2.
In subsection 3.3 we compute in closed form 
the Patterson function within planes 
for vertex points and pentagon centers.
  
The bulk structure of the tiling model, analyzed here up to the level
of Bergman clusters, displays for the planes a repetition and structure
pattern in line with the terrace structure found in 
the experiments \cite{BU} which stimulated the present analysis.
A complementary approach to the terrace structure, based on generating
a model patch, is given in \cite{KR2} and confirms the present
analysis. 

\section{Tilings and windows.}

\subsection{Fibonacci lines, their windows, and search for the string LLSLLSLSLL.}

We recall the well-known projection and window technique for the Fibonacci 
tiling ${\cal T}$. We shall emphasize the window technique since it will
be needed when we apply in 
sections 2.4-2.6 
Fibonacci lines to the icosahedral tiling. Let $\Lambda$ be the square lattice in 2D whose edge length
we adjust for convenience to $\sqrt{\tau+2},\; 
\tau = \frac{1}{2}(1+\sqrt{5})$. In a lattice basis the points
of $\Lambda$ are
\begin{equation}
x = n_1e_1+n_2e_2.
\end{equation}

In a system of coordinates $(x_{\parallel}, x_{\perp})$ rotated by $\phi:\;  c=\cos (\phi)= \frac{\tau}{\sqrt{\tau+2}},
s= \sin (\phi)= \frac{1}{\sqrt{\tau+2}}$ wrt. the natural basis, 
the basis vectors are 
\begin{equation}
e_1 = (c,-s) \sqrt{\tau+2},\, e_2 = (s,c) \sqrt{\tau+2},
\end{equation}
and the coordinates of the lattice points become
\begin{eqnarray}
(x_{\parallel},x_{\perp})
& = & ( n_1e_{1\parallel}+n_2e_{2\parallel},  
n_1e_{1\perp}+n_2e_{2\perp}) 
\\ \nonumber 
& = & (x_{\parallel}(n_1,n_2),x_{\perp}(n_1,n_2))
= (n_1\tau+n_2,-n_1+n_2\tau).
\end{eqnarray}
The projections $(x_{\parallel},x_{\perp})$ form two 
$\tau$-modules on orthogonal lines $E_{\parallel},E_{\perp}$ respectively.
There is a {\em unique map}  
$x_{\parallel}(n_1,n_2) \Leftrightarrow x_{\perp}(n_1,n_2)$ between these 
modules, corresponding to the star map of Moody \cite{MO}, and there is a
{\em unique lifting} of  $x_{\parallel}(n_1,n_2)$ or $x_{\perp}(n_1,n_2)$
into a point of $\Lambda$. The projections 
$x_{\parallel}(\Lambda), x_{\perp}(\Lambda)$ cover $E_{\parallel}, E_{\perp}$
dense and uniformly.

Upon
choosing in $E_{\perp}$ the {\em window} $f_{\perp} := \left(-1,\tau\right]$, 
whose length $|w_{\perp}|= \tau+1$ is the projection of a unit square
to $E_{\perp}$, the vertex set of the Fibonacci tiling ${\cal T}$ 
in $E_{\parallel}$ is defined as
\begin{equation}
v({\cal T}) = \{ x_{\parallel}(n_1,n_2)|  x_{\perp}(n_1,n_2) \in f_{\perp} \}.
\end{equation}
The endpoints of the window are restricted in order to avoid ambiguities.
When ${\cal T}$ is lifted back into $\Lambda \in E^2$, it forms the vertex set of
a continuous staircase formed by edge lines as shown in 
Fig.1.

\begin{center}
\input quadrat
\end{center}

Fig.1 The Fibonacci tiling is the projection of a staircase, formed by 
edge lines in a square
lattice, to a line $E_{\parallel}$ of slope $\tau^{-1}$. 
The successive vertices $x_{\parallel}$ of the 
staircase may be enumerated by the single integer $N=n_1+n_2$.
Projected to $x_{\parallel}$, the edge lines form the Fibonacci tiling
with two tiles S,L of length $1,\tau$ respectively.
The projections $x_{\perp}(N)$ of the vertices to the orthogonal space $E_{\perp}$ 
fall into a window $f_{\perp} = \left(-1,\tau\right]$ of length $\tau+1$. 
\vspace{0.2cm}

The projections of the steps to $E_{\parallel}$ form the familiar 
Fibonacci tiling with two tiles 
S,T of length $1,\tau$ respectively. With $x_{\parallel}$ increasing,
adjacent tiles form the vertex configurations LS, LL, or LS respectively.
The windows in $E_{\perp}$ for these vertex configurations can be shown 
to form
subwindows of $f_{\perp}$ given by
\begin{eqnarray}
f_{\perp}^{LS} & = & \left(-1, 0 \right],
\\ \nonumber
f_{\perp}^{LL} & = & \left(0,\tau-1 \right],
\\ \nonumber
f_{\perp}^{SL} & = & \left(\tau-1,\tau \right].
\end{eqnarray}
We now wish to compare and analyze Fibonacci tilings with different
starting points. Because of the uniform dense covering, we may choose
in $E_{\perp}$ an arbitrary initial point $c_{\perp} \in w_{\perp}$
and associate to it an initial point of a tiling ${\cal T}(c_{\perp})$.
We label the initial vertex by $(0,0)\Rightarrow 0$ and 
the successive vertices of ${\cal T}(c_{\perp})$ by the single
integer $N=n_1+n_2$. From the window condition we can generate
$x_{\parallel}(N),\, x_{\perp}(N)$ step by step according to  
\begin{eqnarray}
x_{\perp}(N+1)
& = & \left[ 
\begin{array}{l}
x_{\perp}(N)-1 \leftrightarrow (x_{\perp}(N)-1) \in f_{\perp},\\  
x_{\perp}(N)+\tau  \leftrightarrow (x_{\perp}(N)+\tau) \in f_{\perp}.\\
\end{array}
\right],
\\ \nonumber
x_{\parallel}(N+1)
& = & \left[ 
\begin{array}{l}
x_{\parallel}(N)+\tau,\\  
x_{\parallel}(N)+1.\\
\end{array}
\right],
\end{eqnarray}  
The steps in  $E_{\parallel}$ propagate the tiling by a
new tile L or S respectively. For later purposes, like the determination of
densities of points in section 3.2, we emphasize 
the propagation as a function of N in terms of the perpendicular 
coordinate in the window.

\begin{center}
\input w3
\end{center}

Fig.2 Four Fibonacci lines starting at an $LL$ vertex are coded
by four initial points in a vertical subwindow scaled by $\tau^{-3}$. 
The vertical coordinate is $y_{\perp}(N)$ of eq.(8).  
For steps numbered from 0 to 24,
the lines connect the images in the window for these four points.
Each step produces in $E_{\parallel}$ a long
or short interval of the corresponding Fibonacci line. 
\vspace{0.2cm}

We adjust
the perpendicular coordinate to the midpoint of the window and scale
it by a factor $\tau$ to obtain the new variable
\begin{equation}
y_{\perp}(N) := \tau x_{\perp}(N)-\frac{1}{2}
\end{equation} 
whose window 
$w_{\perp} = \left( -\frac{1}{2} \tau^3,  \frac{1}{2} \tau^3\right]$ 
now has the length $|w_{\perp}| = \tau^3$ with the central subwindow
for LL vertex configurations of length $|w_{\perp}^{LL}|=1$.
It can be shown that the subset of LL vertices of the original tiling 
form another Fibonacci vertex set scaled by a factor $\tau^3$. 
The function $y_{\perp}(N)$ is plotted in Fig.2 for four initial values
from the subwindow $w_{\perp}^{LL}$. This Figure illustrates the variety of
sequences as a function of the initial value. Successive values are connected by 
straight lines. The reason for starting at an  LL subwindow 
will become apparent when we go to the icosahedral tiling in 
section 2.4.

With the window technique we search for the finite 
string LLSLLSLSLL found in the terrace spacing of the experiment \cite{BU}. 
For a Fibonacci line coded by the initial point 
$y_{\perp}(0)=-\frac{1}{2}$,
this string occurs at the points N= 9 $\ldots$ 19, 
compare Figs. 2,6. For other initial points, the string would occur
at some other step. We infer
all possible occurrences of the string as conditions with respect to the window: 
The string will
be stable under vertical shifts $\Delta y_{\perp}(0)$ of the initial
point as long as
its highest value $y_{\perp}(17)$ and its lowest value 
$y_{\perp}(14)$ do not pass the limits $\pm \frac{1}{2} \tau^3$
respectively of the window $w_{\perp}$. These window conditions are
independent of the initial point.
Clearly the appearance of the
string puts narrow bounds on the corresponding values of $y_{\perp}$,
compare section 2.6.

\subsection{Icosahedral tilings.}

The construction of 3D tilings follows the paradigm given by
the Fibonacci line. The projections are now determined by
requiring non-crystallographic and in particular icosahedral 
point symmetry after projection.
It is well-known that an icosahedral tiling ${\cal T}^{P}$
with two rhombohedral tiles
arises by icosahedral projection to 3D from the primitive hypercubic P-lattice
and module 
in 6D \cite{KR0}. 

In two orthogonal 3D spaces $E_{\parallel},\, E_{\perp}$ 
we find the six 5fold, ten 3fold and fifteen 2fold axes 
associated with the icosahedral group. 
The six primitive 
basis vector $e_1, \ldots e_6$ of the hypercubic lattice 
upon projection point along 5fold axes.
Their length we denote by \ffo, and
their directions we choose as follows
\cite{KR}: In $E_{\parallel}$ we 
take $\cos (e_{1\parallel},e_{i\parallel})= 1/\sqrt{\tau +2},
\,i=2\ldots 6$ and for $i=2 \ldots 5$ pass from 
$e_{i\parallel}$ to $e_{i+1\parallel}$ by a rotation around $e_{1\parallel}$ with angle
$2\pi/5$. In $E_{\perp}$ we 
take $\cos (e_{1\perp},e_{i\perp})= - 1/\sqrt{\tau +2},\,
i=2\ldots 6$ and for $i=2 \ldots 5$ pass from $e_{i\perp}$ to 
$e_{i+1\perp}$ by a rotation around $e_{1\perp}$ with angle
$4\pi/5$. All vectors along 2fold axes arise from projections of 
$(e_i \pm e_j),\, i \neq j$. Their shorter length we denote
by \zfo$\; = \frac{2}{\sqrt{\tau+2}}$\ffo.

The icosahedral quasicrystals i-AlFeCu and
i-AlPdMn from their diffraction pattern are indexed 
by the hypercubic face-centered or F-lattice and -module  rather than 
the primitive
P-module. The hypercubic F-lattice,
scaled by a factor 2 and denoted here as (2F), 
may be viewed as the subset of even lattice
points (even index sum) from the full hypercubic lattice in 6D.

Turning attention to this lattice and module, 
we briefly summarize the construction of the icosahedral
tiling ${\cal T}^{*(2F)}$ associated with the F-lattice
and  given in \cite{KR}. 
In both 3D spaces we have 6D modules whose
bases can be formed for example from three short and three long vectors 
along three selected 2fold axes. By the unique lifting and projection,
there is a one-to-one map
$q_{\parallel} \Leftrightarrow q_{\perp}$, the star map of \cite{MO}, 
between points $q_{\parallel}$ and $q_{\perp}$
of the two   modules in $E_{\parallel}$ and $E_{\perp}$.
For simplicity we suppress the basis and the six integers in
$q_{\parallel}, q_{\perp}$ which generalize eq.(4) and underlie this map. 
The vertex points
of the tiling ${\cal T}^{*(2F)}$ are, as a generalization of 
eq.(5), given by
\begin{equation}
v({\cal T}^{*(2F)}) 
= \{ q_{\parallel} | q_{\perp} \in {\rm triacontahedron} \},
\end{equation} 
i.e. the projections $q_{\parallel} \in E_{\parallel}$ 
of those lattice points whose projections $q_{\perp} \in E_{\perp}$ 
fall into
the  triacontahedral window, compare 
Fig.5. 
The projections $\{ q_{\perp} \}$ 
fill the triacontahedron dense and uniformly.
The triacontahedron is the
icosahedral projection to $E_{\perp}$ of the Voronoi or Wigner-Seitz cell
of the F-lattice in 6D. The tiling ${\cal T}^{*(2F)}$
has six tetrahedral tiles. In its present simple form 
we need only two tetrahedra with
3fold symmetry axis. 
The vertices of these two tetrahedra coincide with four
even vertices of the two rhombohedra associated with the tiling 
${\cal T}^P$. The simple form of the tiling  ${\cal T}^{*(2F)}$
is fully described by putting
atoms into positions on the full rhombohedral tiles but allowing for the distinction
of even and odd vertices, as is done in the Elser model \cite{elser, KR1}.
We shall need only the even and odd vertex points.

The relation of the tilings ${\cal T}^{*(2F)}$
and ${\cal T}^P$ may be summarized as follows: The triacontahedral
windows for the vertex sets coincide. The modules differ from
one another: The (2F) module is the even submodule of the P module. 
By expanding each of the two
3fold symmetric tetrahedra back into
the corresponding rhombohedron and dropping the distinction between 
even and odd vertices we can locally derive ${\cal T}^P$
from  ${\cal T}^{*(2F)}$.

\subsection{Planes perpendicular to 5fold axes.}

We turn to planes in the tiling ${\cal T}^{*(2F)}$.
Fix in $E_{\parallel}$ a 5fold axis parallel to $e_{1\parallel}$ 
as in  Fig.4 and consider
vertex points $q_{\parallel}$ 
in a plane perpendicular to it.
Next we pass to $E_{\perp}$, consider the corresponding 5fold axis
parallel to $e_{1\perp}$ and the images $q_{\perp}$ of the vertex points $q_{\parallel}$ 
from the plane under the one-to-one map.
It turns out that these images $q_{\perp} \in E_{\perp}$ lie again in a plane
perpendicular to the 5fold axis. In addition they must be points
from the triacontahedron. Hence we get the result:

\begin{center}
\psfig{file=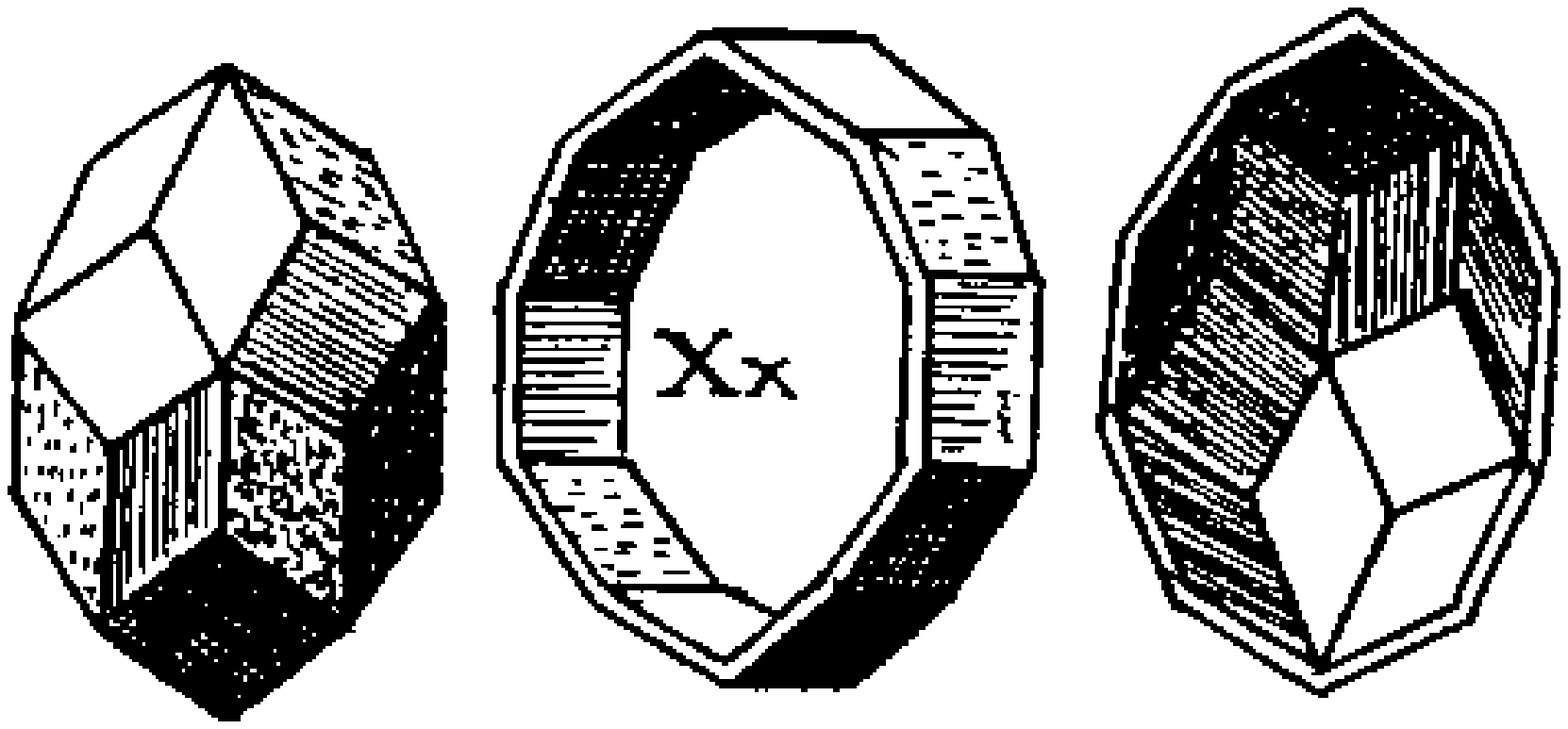, scale=0.7}
\end{center}

Fig.3 Kepler's decomposition of the triacontahedron into a central
decagonal prism $Xx$ and two shells.\vspace{0.5cm}

{\bf Prop 1}: The window for vertex points $q_{\parallel}$ 
from the tiling  ${\cal T}^{*(2F)}$ in a fixed plane perpendicular
to a 5fold axis is in $E_{\perp}$ the intersection of a plane
perpendicular to the corresponding 5fold axis with the triacontahedron.

The triacontahedron is shown in 
Fig.5
in a view perpendicular to a 5fold axis. 
The distance from the center to a 5fold vertex is $\tau$ \ffo,
where \ffo $\,$ is the standard length along a 5fold axis. 
The triacontahedron
with respect to this 5fold axis has a central decagonal prism of thickness
$\frac{2\tau^{-1}}{\tau+2} \tau$ \ffo. J. Kepler in 1619 \cite{KE} not only 
introduced the triacontahedron, but also visualized these
decagonal prisms and denoted them by the letters $Xx$, see Fig.3. 

The central decagonal prism of the triacontahedron when seen as a
subwindow for part of the tiling ${\cal T}^{*(2F)}$ has a particular
siginificance, as is shown in \cite{KR3}: 
Any planar decagonal intersection of the triacontahedron
in $E_{\perp}$ determines in $E_{\parallel}$ an infinite  planar tiling 
TTT by two golden triangles, compare \cite{BAA}. 
With respect to the full 3D tiling 
${\cal T}^{*(2F)}$, this planar subtiling is formed by faces of 
tetrahedral tiles.

The planar intersections of the triacontahedron outside the 
decagonal prism are windows for planes of vertex points which in general
do not form a planar tiling. From the uniform covering of the window
it follows that the density of vertex points in all planes is proportional
to the area of its window, i.e. of the corresponding intersection of
the triacontahedron. We shall compute this density in 
subsection 3.2.

\subsection{Fibonacci shifts between parallel planes.}

The planar TTT subtiling has the property that through any vertex
point there passes at least one infinite Fibonacci line. In terms
of its decagonal subwindow this results from the geometric property 
that any interior point belongs to at least one subwindow for
an infinite Fibonacci line. In $E_{\parallel}$ this Fibonacci line points
along a 2fold axis associated with two vectors whose length 
scales by $\tau$. 
All subwindows for a fixed Fibonacci line are sections of length 
$\tau^2$ \zfo $\;$ on parallel lines perpendicular
to and bounded by opposite rectangular faces of a decagonal
prism.

\begin{center}
\input axes
\end{center}

Fig.4 Two typical perpendicular 2fold axes $2,\, 2'$ can be chosen within a 
plane with a 5fold axis $5$. A plane perpendicular to this axis can be 
shifted from the origin by vectors along these 2fold axes.\vspace{0.2cm}

As the initial plane we shall choose a reference plane whose vertex points
form a triangle TTT pattern. All these planes have
the same highest density of vertex points, see 
subsection 3.2.
To shift between planes perpendicular to a fixed 5fold axis,
we shall use vectors along 2fold axes outside this plane.  
From the orbit in $E_{\parallel}$ of 2fold axis with respect to
the 5fold one we pick two perpendicular 2fold coplanar axes $2, 2'$ 
which form with the  5axis the angles 
$\arccos( \frac{1}{\sqrt{\tau+2}})=58.3$ degrees, 
$\arccos( \frac{\tau}{\sqrt{\tau+2}})=31.7$ degrees,
see Fig. 4.
In the notation of \cite{KR}, we choose the axis $5$ along
$e_{1\parallel}$, the axis $2$ along the short and long vectors 
$-(e_2+e_3)_{\parallel},(e_1+e_5)_{\parallel}$, and 
the axis $2'$ along the short and long vectors 
$(e_1+e_5)_{\parallel},-(e_4+e_6)_{\parallel}$ respectively.

The vectors along the 2fold axis $2$ from eq.(1)  have the 
$\tau$-scaled model length 
$\tau$ \zfo, $\tau^2$ \zfo $\;$. 
By multiplication with the cosine of
the corresponding angle we get the 
parallel spacings of
planes perpendicular to the 5fold axis $5$.  These spacings become
($\frac{2\tau}{\tau+2}$ \ffo) $\;$ and ($\frac{2\tau^2}{\tau+2}$ \ffo) $\;$
respectively. By comparison, the two vectors along the 2fold axis $2'$ yield along
the 5fold axis spacings scaled by a factor $\tau$. Therefore the latter
vectors will not generate additional parallel planes. We 
obtain the short and long spacings 4.08 and 6.60 \AA $\;$ respectively, 
fully in line
with the terrace spacing observed in \cite{BU} and quoted in
section 1. So we have identified 
in the bulk tiling model the shift vectors which generate the terrace
structure.

\begin{center}
\input guertel
\end{center}

Fig. 5 Three decagonal prisms of the triacontahedron in $E_{\perp}$:
The first one has its 5fold axis 5 in the vertical direction. 
The second and third prism have opposite rectangular faces 
perpendicular to two 2fold axes 2,2' coplanar with the axis 5.
\vspace{0.2cm}

To assure that, starting from a fixed plane, we generate by shifts
an infinite system  of parallel planes, we turn to $E_{\perp}$.
In $E_{\perp}$ the three axes $2,2',5$ remain coplanar but the directions and
angles of $2$ and $2'$ are interchanged, see Fig.5.
We get infinite Fibonacci lines along the axes 2,2' if the vectors 
along these axes can be associated to decagonal prisms of the
triacontahedron. Two decagonal prisms with this property and associated with
2,2' respectively are shown in Fig.5.

The initial reference plane was chosen with the triangle tiling and hence has as
its window a decagonal section through the triacontahedron perpendicular
to the 5fold axis. An infinite system of parallel planes will arise,
provided that we select a starting point which also belongs to a
Fibonacci window associated with vectors say along the axis 2.
The projection of the full  Fibonacci window along the 5fold axis 
in $E_{\perp}$ by multiplication with $\cos(2,5) = \frac{1}{\sqrt{\tau+2}}$
becomes $\frac{2\tau^3}{\tau+2}$\ffo. The projection of its central subwindow
for LL vertices equals the thickness  $\frac{2}{\tau+2}$\ffo$\,$ of the 
decagonal prism. We conclude that, among the parallel planes
shifted along the infinite Fibonacci line in the direction 2, 
the reference plane and in fact any 
dense plane occurs at the LL vertices. 
\vspace{0.2cm}

We summarize the {\em information obtained so far on parallel planes 
of vertex points} $q_{\parallel}$ {\em in the tiling} ${\cal T}^{*(2F)}$,
generated by Fibonacci lines along 2fold axes: Starting from a dense 
reference plane,
we generate an infinite set of parallel planes. They follow a Fibonacci
spacing with perpendicular short and long distances 
$\frac{2\tau}{\tau+2}$ \ffo $\;$= 4.08 \AA $\;$
and $\frac{2\tau^2}{\tau+2}$ \ffo $\;$= 6.60 \AA, fully in line with the STM
observations of terraces. The dense planes occur at all 
LL vertices of the generating Fibonacci line. Other parallel
planes in the set will have a lower density of vertex points.
The string LLSLLSLSLL analyzed in section 2.1 can now be 
converted into a sequence of parallel planes of varying 
density.

To complete the analysis of parallel planes, we must find out
what fraction of all vertex points $q_{\parallel} \in
{\cal T}^{*(2F)}$ is reached within this infinite sequence
of planes.  

\subsection{Parallel planes are connected by Fibonacci lines.}

We wish to show that indeed we can reach from a fixed dense 
reference plane most vertex points $q_{\parallel}$  
by shifts along Fibonacci lines. For this purpose we consider only
those vertex points $q_{\parallel}$ which lie on at least one infinite
Fibonacci line. From the window side we know that this is 
the case if $q_{\perp}$ is a point from any  decagonal prism. 
There are six such prisms,
and this motivates the definition of a new window:
 
{\bf Def 2}: Decagonal prism $Xx$ approximation: We analyse only those points
$q_{\perp}$ of the triacontahedral window  which belong to at least one decagonal 
prism, that is to the union $\cup_j^6 prism_j$. We omit
in this approximation the points $q_{\perp}$ from 
small parts of the triacontahedron 
close to its 5fold vertices, compare Fig.5.

The vertex points $q_{\parallel}$, with $q_{\perp}$ 
belonging to this new window,  have the following 
properties:

{\bf Prop 3}: Consider vertex points $q_{\parallel}$ in a plane parallel to a 
fixed dense $TTT$ infinite reference plane. 
Among them there is a point on 
an infinite
Fibonacci line which intersects (as a continuous line) the reference plane.

{\em Proof}: Through any point coded in the decagonal prism approximation 
there runs at least one infinite Fibonacci line. If it intersects the 
reference plane we are through. If it runs parallel to the reference
plane, we can (proof omitted) in at most two parallel steps pass 
to another point with an intersecting Fibonacci line.\vspace{0.2cm}

{\bf Prop 4}: If an infinite  Fibonacci line intersects 
as a continous line a $TTT$ plane, it 
hits this plane in a vertex  point.

{\em Proof}: The points of the infinite reference plane form the vertices of the 
planar $TTT$ subtiling by faces, the points  of the  non-parallel infinite 
Fibonacci line form the vertices of a linear subtiling by edges 
of the $3D$ tiling ${\cal T}^{*(2F)}$. Both subtilings are parts, 
hence their intersection is a vertex of the full tiling.
\vspace{0.2cm}

The two propositions allow us to code in $E_{\perp}$  planes of 
vertex points, parallel 
in $E_{\parallel}$ to a first
dense reference plane perpendicular to $5$, by their intersections 
with Fibonacci lines along the axis $2$ of Fig.4:

{\bf Prop 5}: Any vertex plane perpendicular to a 5fold axis 
has at least one point 
connected to the 
reference plane by an infinite Fibonacci line. Conversely,  
by following all non-parallel Fibonacci lines
from the reference plane we reach  any  parallel vertex plane. 
\vspace{0.2cm}

We have shown that all vertex points $q_{\parallel}$ such that
$q_{\perp}$ belongs to the Kepler Xx model appear in an infinite 
sequence  of parallel planes in the order  and spacing of 
a Fibonacci line. The analysis of the 1D Fibonacci system 
given in subsection 2.1 now applies to the 3D tiling. 
The stepwise generation  in $E_{\perp}$ of the 1D Fibonacci
system shown in Fig. 2 can now be converted into a stepwise generation in
$E_{\perp}$ of parallel planes, enumerated by the integer N: Starting at an LL vertex means
starting at a dense plane. The value of the perpendicular coordinate
$y_{\perp}$ yields the value of a coordinate 
$\eta \tau$ \ffo, $-1 \leq \eta \leq 1$ from the center of 
the triacontahedron along the 5fold 
axis. The explicit relation is
\begin{equation}
\eta(N)  = \frac{2\tau^{-1}}{\tau+2} y_{\perp}(N).
\end{equation} 
In Fig.6 we plot the projection of the triacontahedron together with 
the values of $y_{\perp}(N)$ connected by lines. Each value determines 
the height of a 
corresponding horizontal section of the triacontahedron. 
Since the point density in the corresponding plane of the tiling
is proportional to the area of this section, Fig.6 provides 
insight into the variation of this density from step to step.

Note that
within the Fibonacci window we do not reach the highest absolute 
values of $\eta$. For the additional points in Fig.6 see section 
2.6.

\begin{center}
\input w42
\end{center}

Fig.6  Vertical values of  $y_{\perp}(N)  \sim \eta(N)$ eq.(10)
determine horizontal sections in $E_{\perp}$ of the  
triacontahedron as windows for planes. The values are connected by lines 
and follow the 
Fibonacci coding. The total width of the vertical window is 
$\frac{2\tau^2}{\tau+2} \tau$ \ffo. The  triacontahedron has the 
vertical diameter $2\tau$ \ffo. Numbers in the second row assign 
planes corresponding to the terraces
found in \cite{BU}. The bars $1^-,9^-,3^+,6^+,11^+$  
mark values of $\eta$ for 
additional low-density planes of vertex points.

\subsection{From planes to terraces at the surface.} 

We have found in section 2.4 from the bulk model 
sequences of planes with a spacing
that agrees with the terrace spacing found in \cite{BU}.
We now interpret terraces at the surface as particular 
planar terminations  from the bulk tiling model. From Fig.2 we have already 
identified a string in correspondence to the observed string
of terraces. With the numbers $1 \ldots 11$ in the second 
horizontal row of Fig.6 we now 
assign planes and values $\eta_1$ to eleven planes 
which correspond to  the  spacing of eleven high or low terraces
found in \cite{BU}. The numbers follow the terraces
in a direction into the bulk material.

The numbers $\eta_1$ given in Table 1 are not unique, 
but the appearance  of the finite
string puts narrow
bounds on their range:   
Maximal shifts upwards by $\Delta \eta = \frac{2\tau^{-1}}{\tau+2}(2\tau-3)$ 
from $N=9$ or downwards by 
$\Delta \eta = \frac{2\tau^{-1}}{\tau+2}(-3\tau+5)$ from $N=14$ are 
compatible with the appearance of the string, but of course give different values of
the density and of the Patterson function.  
 
As in the patch analysis given in \cite{KR2}, there appear additional
vertex planes in $E_{\parallel}$ with a spacing scaled by $\tau^{-1}$
which do not yield terraces in the experiment. This narrow spacing
cannot be coded by a single Fibonacci line. With respect to the 
triacontahedron in Fig.5, it  requires a vertical shift
$\Delta \eta = \frac{2\tau^2}{\tau+2}$ in $E_{\perp}$. 
This shift can be produced by the sum of two vectors
of length $\tau$\zfo $\;$pointing along the axes $2$ and $2'$ in Fig.2 
respectively. 
The summed vector (not parallel to the
5fold axis) connects points in the 
triacontahedron only if the initial point obeys
$\frac{\tau}{\tau+2} \leq |\eta| \leq \frac{\tau^2}{\tau+2}$.
In the selection of points of Fig.6 the values $\eta$ of the 
corresponding 5 final points
are denoted by $1^{-},3^{+},6^{+},9^{-},11^{+}$.
The $\pm$ sign codes  an additional  vertex plane 
shifted in $E_{\parallel}$ by 
$\frac{2\tau^{-1}}{\tau+2}$ in units $\tau$ \ffo $\;$above or below the 
plane with the fixed number. The positions of these planes agree
with the patch analysis of \cite{KR2}. All these additional 
planes have a very low 
density of vertex points.

In section 3.2 we compute the exact model density of vertex
points in the planes which decreases with the absolute value of 
$\eta_1$. From Fig.6 it can be seen that the planes $3,6,9$ in the 
string have the highest values of $|\eta_1|$ and hence the 
lowest density. In the experiment \cite{BU} these planes 
correspond to terraces of minimum measured planar size.

\section{Clusters, pentagonal faces and cuts,  and correlations in planes.}

We proceed to an analysis of the more detailed model structure within 
the planes.
So far we looked at planes occupied by vertex points from the tiling.
The full set of atomic positions \cite{elser, KR1} comprises
more points in various classes.
The repetition pattern and variation of the density for parallel planes 
found in section 2 is a general
property of the tiling and applies to any set of atomic positions
within planes perpendicular to a 5fold axis. If atomic
positions of two different types appear within the same initial 
fixed plane, their repetition pattern and variation of density 
follows the same pattern as found for vertex points,  
but may propagate from different initial conditions.
The relative density and the correlation of different types of 
atomic positions sitting within
the same plane will then show {\em systematic variations} along a sequence
of planes. 
\vspace{0.2cm}

We consider in this 
section additional atomic
positions from the Bergman clusters.
The vertex positions of Bergman clusters on the tiling will produce points in parallel
planes.
This holds true in particular for top faces of Bergman 
clusters which run perpendicular to the chosen 5fold axis. They
are of particular interest as candidates for the pentagons 
found in \cite{BU}. 
Upon comparing
the height of $6.60 $ \AA $\;$for the Bergman clusters with the lowest terrace
spacing of $4.08 $ \AA $\;$ we already conclude that these clusters are 
{\em cut at  quasicrystal surfaces}.    
A second larger type of pentagons arises from
a top cut at the heigth of  $4.08 $ \AA $\;$ 
through five vertices of the Bergman dodecahedra.
In the full model
\cite{elser, KR1} of $AlPdMn$ these top face 
pentagons have central atoms in  a lowered central position. Therefore
these pentagon faces would produce at their 
centers the observed holes in the terraces \cite{BU}. 
We shall examine the correlation of
both types of pentagons with vertex points of the tiling.

\subsection{Three models for correlations of pentagons and vertex points.}

We analyze three significant types of
planes perpendicular to a 5fold axis taken from the 
bulk model and explore the correlation within these planes.
\vspace{0.2cm}

(i) In a {\em first model analysis} the  vertex points of the tiling
dominate the sequence of perpendicular planes, characterized by the values
$\eta_1$ as before. 

Consider the relation of the Bergman top pentagons with respect to these planes.
As mentioned in the introduction,
the centers of the Bergman
dodecahedra take the positions of those (odd) points of  the primitive
$P$-lattice which are dropped when going from P to (2F). This has the 
consequence that the Bergman centers can be grouped into 
shifted planes perpendicular to a 5fold axis. The planar densities 
of Bergman clusters in these planes must follow the same rules,
minima, and maxima
as  given for vertex positions in sections 2.4 and 3.2.

The 
height of the Bergman dodecahedra is such that a 5fold vector
from the plane to the center, passing  through a pentagonal face,
has the length and direction of a typical vector $e_{5\parallel}$,
compare Fig.7.
With the chosen enumeration of terraces we follow them 
in the direction of the vector $e_{1\parallel}$,
{\em in the direction} of the 5fold axis in Fig.4. 
Now we look for Bergman
clusters with a center displaced below and a top pentagon  
within a fixed plane of vertex points.
\begin{center}
\input dodek
\end{center}

Fig.7 A Bergman dodecahedron touches with a top pentagon 
from below a plane of vertex points marked by the top horizontal line. 
The vector $e_{1\parallel}$ points {\em downwards}. 
A typical vector downwards 
from the plane  
to the Bergman center is $e_{5\parallel}$. 
A vector like 
$(e_1-e_5)_{\parallel}-e_{5\parallel}$ runs
from a planar pentagonal top cut through the 
dodecahedron, marked by the lower horizontal line, 
to the Bergman center.\vspace{1cm}
 
Transforming the vector $e_{5\parallel}$
from $E_{\parallel}$ to  $E_{\perp}$, compare section 2.2, 
one finds the coding points for
all centers of the Bergman dodecahedra displaced
by $e_{5\perp}$
{\em against the direction of the axis} $5$ of Fig.5, hence downwards
in Fig.5, from the plane coding the vertex points.
So the planes with a corresponding {\em downwards shift} 
$\Delta \eta = \frac{1}{\tau+2}$ form another Fibonacci sequence that  
codes the Bergman centers. The shifted values $\eta_2$ for this sequence 
are given in the second column of Table 1. In Fig.8 we show the two values
$\eta_1,\eta_2$ as functions of $N=0,\ldots, 24$. If the vertical shift 
from the crosses to the circles equals 
$\Delta \eta = \frac{1}{\tau+2}$, the 
Bergman top pentagons touch the vertex plane from below. 
This holds true except for the planes $N=6, 11, 14, 19$, see Fig.8.
In the Fibonacci sequence of planes of vertex points this
correspond to an LS vertex, but in the planes of Bergman faces 
to an SL vertex. Therefore no Bergman top faces can appear in these
planes of vertex points. In the string of planes $9 \ldots 19 \rightarrow
1 \ldots 11$ this would occur at the planes 
$11, 14, 19  \rightarrow 3, 6, 11$. These
planes carry instead the larger Bergman top cut pentagons,
see model (iii).

Figs.8,9 demonstrate that planes formed from atomic positions
of different type (vertex points or pentagon centers) follow the same 
Fibonacci propagation law. Due to systematic shifts in the parameter
$\eta$ for different types, the corresponding densities, which are functions
of this parameter, propagate differently even for atomic positions 
of different type within the same initial plane.

\newpage
\begin{center}
\input treppe1
\end{center}

Fig.8. The values $y_{\perp}(N) \sim \eta(N)$ 
for $N=0,\ldots, 24$ in model (i) determine horizontal sections of the 
triacontahedron as windows for planes of vertex points ($\eta_1$, crosses) 
or of centers for Bergman top pentagons ($\eta_2$, circles).

\begin{center}
\input treppe2
\end{center}

Fig.9. The values $y_{\perp}(N) \sim \eta(N)$ 
for $N=0,\ldots, 24$ in model (ii) determine horizontal sections of the
triacontahedron as windows for planes of centers for Bergman top 
pentagons ($\eta_1$, crosses)
or of vertex points ($\eta_3$, circles).

(ii) In the  {\em second model analysis} of the planes determined by $\eta_1$
we assume  that they are dominated by 
top pentagons of Bergman dodecahedra located below the plane, compare Fig.9. 
The additional presence
of vertex points on these planes is now coded by a 
{\em positive shift} $\Delta \eta = \frac{1}{\tau+2}$. In Table 1 we give the
new values $\eta_3$ for these vertex planes and show them in
Fig.9. Additional vertex points appear except in  
the selected planes $9, 17  \rightarrow 1, 9$. 
\vspace{0.2cm}

(iii) In the {\em third model analysis} we assume that the 
planes determined by $\eta_1$ are dominated
by pentagons, scaled by $\tau$ and corresponding to top cuts through 
Bergman
dodecahedra below the plane. A typical vector from the plane for this cut to
the center of the Bergman cluster in the notation of section 2.2  
is $(-e_{5\parallel}+e_{1\parallel})-e_{5\parallel}$ indicated 
in Fig.7.  
The edge size \zfo $\; = 4.78$ \AA $\;$of the pentagons would be in line with
the  holes observed in \cite{BU}, and it would also lead to 
a central hole. 
Again we ask about the presence of additional vertex
points. From the relative position of these pentagons to the vertex
planes as shown in Fig.7 we deduce in $E_{\perp}$ 
a shift $\Delta \eta = \frac{2\tau+1}{\tau+2}$. At $N=0$ this shift modulo the 
window size becomes $\frac{-1}{\tau+2}$ which then generates  the values 
$\eta_3$ given in Table 1  and shown in Fig.9. Vertex points can
occur only  if the relative shift from crosses  to circles  is 
$\Delta \eta = \eta_3-\eta_1=\frac{2\tau+1}{\tau+2}$. This happens
only in the selected planes $9, 17 \rightarrow 1, 9$, otherwise  
there are no vertex points within these planes. A further shift analysis 
shows that any plane containing Bergman top pentagons cannot contain   
Bergman cut pentagons and vice versa. In model (iii) there appear
Bergman top cut pentagons in densest planes. A closer inspection shows 
that these pentagons share vertices  and form an almost 
connected graph.  

The three model cases considered yield three interpretations of bulk planes 
as terminations  for the terraces 
observed in \cite{BU}, all in line with the relative spacing. 
In cases (i,ii) the planes carry vertex points or  
pentagons corresponding to 
the faces of Bergmann dodecahedra. In case (iii) the planes carry 
the larger cut pentagons and almost no vertex points.

\subsection{Planar density of atomic positions.}

We compute the 
exact area $F(\eta)$ of a planar section of the triacontahedron as 
a function of $\eta, -1 \leq \eta \leq 1$. This function is proportional to the exact
density of vertex points $D(\eta)$ in the plane coded by this section. The result is
\begin{eqnarray}
0 \leq & |\eta| & \leq \frac{\tau^{-1}}{\tau +2}:
\\ \nonumber
F(\eta) & = & (\tau+2)^{-3/2} \left[ 10 \tau\right],
\\ \nonumber
\frac{\tau^{-1}}{\tau +2} \leq & |\eta| & \leq \frac{\tau}{\tau +2}:
\\ \nonumber
F(\eta) & = & (\tau+2)^{-3/2} \left[ 10 \tau 
-5\frac{(\tau+2)^2}{\tau}(|\eta|-\frac{\tau^{-1}}{\tau+2})^2 \right],
\\ \nonumber
\frac{\tau}{\tau +2} \leq & |\eta| & \leq \frac{\tau^2}{\tau +2}:
\\ \nonumber
F(\eta) & = & (\tau+2)^{-3/2} 
\\ \nonumber
& & \left[ 10 
+5\frac{(\tau+2)^2}{\tau}(\frac{\tau^2}{\tau+2}-|\eta|)^2 
-5(\tau+2)^2(|\eta|-\frac{\tau}{\tau+2})^2 \right],
\\ \nonumber
\frac{\tau^2}{\tau +2} \leq & |\eta| & \leq 1:
\\ \nonumber
F(\eta) & = & (\tau+2)^{-3/2} 
\left[ 5 (\tau+2)^2(1-|\eta|)^2 \right]. 
\end{eqnarray}
The function $F(\eta)$ is plotted in Fig.10. The maximum is $F(0)= 10\tau 
(\tau+2)^{-3/2} = 2.3511 $, the maximum value of $\eta$ for the Fibonacci
sequence of planes is 
\begin{equation} 
|\eta|=\tau^2(\tau+2)^{-1}= 0.7236.
\end{equation}
These values are marked by vertical and horizontal lines in Fig.10.

The density can be converted into the 
absolute density of vertex points  by considering the 
densest planes with a triangle pattern and its vertices: 
In the triangle  pattern, each triangle
contributes, because of the sum $\frac{1}{2}\, 2\pi$ of its angles, 
a weight $\frac{1}{2}$ to the number of vertex points.
In terms of the short edge length $s$, the area $f_1,f_2$ and relative 
frequency $\nu_1,\nu_2$ of the large and small triangle are 
\begin{equation}
f_1=s^2 \frac{\tau}{4} \sqrt{\tau+2},\;
f_2= \tau^{-1}f_1,\; \nu_2 = \tau^{-1} \nu_1,
\end{equation}
These expressions yield for the {\em density of vertex points}
(number of vertices per unit area) the value 
\begin{equation}
D(0)= \frac{1}{2f_1}\frac{\tau^3}{\tau+2}
= s^{-2} \frac{2\tau^2}{{(\tau+2)}^{3/2}} 
\end{equation}
In the present model we put for the short edge length 
$s= \tau$ \zfo $\; = \tau \frac{2}{\sqrt{\tau+2}}$ \ffo $\;$and 
\ffo $\; = 4.56$ \AA $\;$ eq.(1) to obtain the model value
\begin{equation}
D_{max,(i)} = D(0)= 12.6 \cdot 10^{-3} \mbox{\AA}^{-2}.
\end{equation}
The density of vertex points in a plane for fixed $\eta$ is now
computed as  
\begin{equation}
D(\eta) = D(0) \frac{F(\eta)}{F(0)}.
\end{equation} 
The lowest density in a Fibonacci sequence of planes from eqs.(14,15) 
is 
\begin{equation}
D_{min}= \frac{1}{2\tau} D(0) = 0.3090 D(0).
\end{equation}
These values of the density refer to vertex positions in the tiling.
All other atomic positions on the tiles will of course lead to
other specific atomic densities and correlations.

The maximum density of vertex points, associated with model (i),
is given in eq. (16). The  same maximal density applies to the 
centers of Bergman top faces in model (ii). Each pentagon contributes 
$5$ vertex positions which would yield a factor $5$ for
the pentagon vertex density. For model (iii) with the larger pentagons,
the maximum density of centers is still the same. For the density of
pentagon vertices one should not multiply by a factor 5: It turns out
that these larger pentagons in a dense plane can share vertices. An exact 
computation of the maximum density for the vertices of large pentagons
yields (proof omitted) 
\begin{equation}
D_{max,(iii)}
= \frac{7\tau+4}{\tau^3} D_{max,i}= 3.6180 D_{max,(i)}. 
\end{equation}

From the density we can compute the average distance by comparison
for example with a tiling by equilateral triangles. For such a tiling,
the density of points is related to the edge length t by 
\begin{equation}
D_3(t) = t^{-2} \frac{2}{\sqrt{3}}
\end{equation}
If we equate this density with the expression eq.(15) found in the 
dense vertex planes, we obtain an equivalent distance
$t_{eq} = 9.5 $ \AA $\;$.

In columns 5-7 of Table 1 we give  $F$ 
for the three values of $\eta$. Note that, in view of
the three models discussed in subsection 3.1 , the three values of $F$
in a row do not always refer to the same plane.

\newpage
\begin{center}
\input approxi
\end{center}

Fig.10 The function $F(|\eta|)$ for a plane  fixed by $\eta \geq 0$ 
is in $E_{\perp}$ the area of a planar section of the triacontahedron.
The horizontal line marks the lowest value of $F$, the vertical line 
the highest value of $|\eta|$ in a Fibonacci sequence of planes. 
In $E_{\parallel}$, $F(|\eta|)$ is
proportional to the 
density of points or pentagon centers within a plane.
\vspace{0.5cm}

\newpage
{\bf Table 1}:
The values $\eta_1,\eta_2,\eta_3$ as functions of $N=0,\ldots,24$
code inside the triacontahedron 
Fibonacci sequences  of planes
perpendicular to a 5fold axis. These values are associated 
alternatively with vertex
points, pentagonal faces and cuts of Bergman dodecahedra.
Columns $5-7$ give the value of $F(\eta_i)$, equal to 
the area of planar sections through the triacontahedron and proportional to the 
relative density of points. The rows $9 \ldots 19$ are put in
correspondence with the terraces $1 \ldots 11$ of \cite{BU}.

\begin{center}
\begin{tabular}{|r|r|r|r|r|r|r|} \hline
  N & $\eta_1$&$\eta_2$ &$\eta_3$ &$F(\eta_1)$&$F(\eta_2)$&$F(\eta_3)$ \\ \hline
  0 & -0.1708 & -0.4472 &  0.1056 &  2.3511 &  1.9021 &  2.3511 \\ \hline
  1 &  0.7236 &  0.4472 & -0.4472 &  0.7265 &  1.9021 &  1.9021 \\ \hline
  2 &  0.1708 & -0.1056 &  0.4472 &  2.3511 &  2.3511 &  1.9021 \\ \hline
  3 & -0.3820 & -0.6584 & -0.1056 &  2.0891 &  1.0541 &  2.3511 \\ \hline
  4 &  0.5125 &  0.2361 & -0.6584 &  1.6746 &  2.3261 &  1.0541 \\ \hline
  5 & -0.0403 & -0.3167 &  0.2361 &  2.3511 &  2.2260 &  2.3261 \\ \hline
  6 & -0.5931 &  0.5777 & -0.3167 &  1.3507 &  1.4162 &  2.2260 \\ \hline
  7 &  0.3013 &  0.0249 &  0.5777 &  2.2510 &  2.3511 &  1.4162 \\ \hline
  8 & -0.2515 & -0.5279 &  0.0249 &  2.3129 &  1.6164 &  2.3511 \\ \hline
  9 &  0.6430 &  0.3666 & -0.5279 &  1.1269 &  2.1259 &  1.6164 \\ \hline
 10 &  0.0902 & -0.1862 &  0.3666 &  2.3511 &  2.3497 &  2.1259 \\ \hline
 11 & -0.4626 &  0.7082 & -0.1862 &  1.8512 &  0.8067 &  2.3497 \\ \hline
 12 &  0.4318 &  0.1554 &  0.7082 &  1.9508 &  2.3511 &  0.8067 \\ \hline
 13 & -0.1210 & -0.3974 &  0.1554 &  2.3511 &  2.0495 &  2.3511 \\ \hline
 14 & -0.6738 &  0.4971 & -0.3974 &  0.9796 &  1.7311 &  2.0495 \\ \hline
 15 &  0.2207 & -0.0557 &  0.4971 &  2.3365 &  2.3511 &  1.7311 \\ \hline
 16 & -0.3321 & -0.6085 & -0.0557 &  2.1982 &  1.2835 &  2.3511 \\ \hline
 17 &  0.5623 &  0.2859 & -0.6085 &  1.4800 &  2.2733 &  1.2835 \\ \hline
 18 &  0.0095 & -0.2669 &  0.2859 &  2.3511 &  2.2969 &  2.2733 \\ \hline
 19 & -0.5433 &  0.6276 & -0.2669 &  1.5565 &  1.1980 &  2.2969 \\ \hline
 20 &  0.3512 &  0.0748 &  0.6276 &  2.1600 &  2.3511 &  1.1980 \\ \hline
 21 & -0.2016 & -0.4780 &  0.0748 &  2.3456 &  1.7986 &  2.3511 \\ \hline
 22 &  0.6928 &  0.4164 & -0.4780 &  0.8851 &  1.9966 &  1.7986 \\ \hline
 23 &  0.1400 & -0.1364 &  0.4164 &  2.3511 &  2.3511 &  1.9966 \\ \hline
 24 & -0.4128 & -0.6892 & -0.1364 &  2.0070 &  0.9033 &  2.3511 \\ \hline
\end{tabular}
\end{center}
\vspace{0.2cm}

The three models (i,ii,iii) given in subsection 3.1 yield different
density values $F$. In model (i), the density of vertex points and 
centers of Bergman top faces is given by $F(\eta_1)$ and,
with exceptions, by $F(\eta_2)$. In the plane $16 \rightarrow 8$,
the density of Bergman faces becomes $D = 6.8\; 10^{-3}$ \AA$^{-2}$.
respectively. In model (ii), the density of Bergman top faces and 
vertex points is given by  $F(\eta_1)$ and with exceptions
by $F(\eta_3)$.
respectively. In model (iii), the density of Bergman top cuts is 
again given
by $F(\eta_1)$.

The density of pentagonal holes in the experimental data \cite{BU} 
has approximately \cite{KR2} the  
value $D= 4.2\; 10^{-3}$ \AA$^{-2}$.
This value would favour model (i) with Bergman top faces in planes
dominated by vertex points.

\subsection{Patterson analysis in planes.}

Consider the general Patterson function $P(x_{\parallel})$ 
of a quasiperiodic point set with a window $W$ at each point of the
lattice $\Lambda$. Let $v_{\parallel }$ be 
a shift vector projected from a lattice vector $v$. 
Let $\chi(x_{\perp})$ 
be the characteristic function of the window $W$ for the lattice points 
projected to $E_{\perp}$. 

{\bf Prop 6}: The Patterson function in $E_{\parallel}$ 
at  $x_{\parallel}= x-x_{\perp}$ is given
by
\begin{equation}
P^{total} (x_{\parallel})= 
\sum_{v \in \Lambda} \delta(x_{\parallel}-v_{\parallel})
\int_{W} \chi(x_{\perp})\chi(x_{\perp}-v_{\perp}) dx_{\perp}
=\sum_{v \in \Lambda} \delta(x_{\parallel}-v_{\parallel}) P(v_{\perp}). 
\end{equation}  

We shall put $P(v_{\parallel})=P(v_{\perp})$.
A Patterson analysis within a plane perpendicular to the 5fold axis $5$
with a fixed shift vector $v_{\parallel}$ parallel to this plane
involves the following notions
in $E_{\perp}$:
Consider the corresponding planar section of the triacontahedron 
with an intersection at $\eta \; \tau$ \ffo $\;$along the 5fold axis $5$.
The values of $\eta$ for the $11$ selected planes are given in
Table 1 both for the vertex points and for the centers of Bergman
dodecahedra touching the vertex plane from below. The area of the 
planar section is proportional to the density $D(\eta)$ of vertex points.
The terraces 2,5,10 in this interpretation
are dense vertex planes. The terraces 2 and 7 have the highest,
the terrace 6 has the lowest density  of Bergman faces. 

The Patterson function from Prop.4 is computed as follows: Shift the planar
section by the vector $v_{\perp}$ parallel to itself.
Compute the area of the intersection of the shifted and unshifted 
section. This area is proportional to the value of the Patterson 
function in $E_{\parallel}$ at the point $v_{\parallel}$. 
The value of $P(0)$ is proportional to the 
density of points. 
We may also normalize by plotting $P(v_{\parallel})/P(0)$. \vspace{1cm}

In a {\em circle approximation} we proceed as follows: 

We use the 
exact area $F(\eta)$ of a planar section of the triacontahedron as 
a function of $\eta$. 

For any fixed value of $\eta$ we compute the radius $r(\eta)$
of a circle with the same area as the section, hence
\begin{equation}
r(\eta) := \sqrt{\pi^{-1} F(\eta)}.
\end{equation}
This amounts to replacing the triacontahedron by a rotational surface
whose circular areas (and hence density values)  for any $\eta$ are 
equal to 
the ones  of the triacontahedron. 
Then we approximate the Patterson function by using these
circles instead of the polygonal sections as functions of $\eta$.
In this approximation, the Patterson function only depends on $|v_{\perp}|$.

The Patterson function $P$ has a smooth behaviour 
in terms of the two variables $\eta$ and
$v_{\perp}$. This expression is shown in Fig.11. 
For $v_{\perp}=0$ it reduces to the function $F(\eta)$ 
and for $v_{\perp}=2r(0)$ it must go to  zero.
 
\begin{center}
\input d3plot
\end{center}

Fig.11 The Patterson function $P(\eta,v_{\perp})$.
For $v_{\perp}=0$ it reduces to $F(\eta)$.
\vspace{0.2cm}

For the plane number $16 \rightarrow 8$, we give in Figs. 12-14 the values of the
Patterson function $P$ for $\eta_1,\eta_2, \eta_3$, represented by 
areas of circles, as a function of the $11$ points
$v_{\parallel}$ selected in a plane as in \cite{BU}.

\newpage
\begin{center}
\input auton1
\end{center}

\begin{center}
\input auton2
\end{center}

\begin{center}
\input auton3
\end{center}

Figs.12-14. The Patterson function $P$ corresponding to the 
values $N=16 \rightarrow 8$ for the values 
$\eta_1,\eta_2,\eta_3$ from Table 1, represented by the
area of circles in a plane. The upper right-hand circle 
stands for $0: v_{\parallel}=0$, the first point $I'$ to its left
is at a distance $|v_{\parallel}|= 7.78 $ \AA $\;$
the others are labelled by roman numbers 
$I \ldots X$ in correspondence to \cite{BU}.\vspace{0.2cm}

In Figs.15-17 we present the same values as a function of
the 10 roman numbers which label peak positions $v_{\parallel}$. 
It can be seen that 
$P(\eta_2)$ yields the lowest density and the 
strongest relative variation, $P(\eta_1)$ and $P(\eta_3)$ are 
very similar, but the latter yields the highest density.

\begin{center}
\input berggflaabs
\end{center}

\begin{center}
\input bergkflaabs
\end{center}

\begin{center}
\input gittflaabs
\end{center}

Figs.15-17. Patterson function $P$ as in the previous
Figure as a function of the roman indices
$I',I,\ldots X$ for the three values $\eta_1,\eta_2,\eta_3$
from top to bottom.
\vspace{0.2cm}

We can now compare the 
three different models (i,ii,iii) of the terrace structure 
given in subsection 3.1  in terms of the Patterson data for terrace 8. 
Model (i) yields the lowest value of the Patterson data for the 
Bergman top faces, along with a larger value for the vertex
points. Models (ii) and (iii) give the same larger values of
the Patterson function for the small or the large top pentagons, 
but differ in the presence or absence of vertex points respectively.
All three models are compatible with the qualitative set of 
experimental Patterson data \cite{BU}.

\section{Conclusion.}

We analyze  a tiling model for the surface structure of 
i-$Al_{68}Pd_{23}Mn_9$ quasicrystals perpendicular to a 5fold axis.
The surfaces are interpreted as terminations in atomic planes 
from the bulk tiling model
decorated by Bergman clusters. The model is analyzed 
in a window  approach. 
The quasiperiodic relations between planes occupied by atoms are 
rigorously and quantitatively derived 
from the bulk model. Sequences of planes are shown to be generated and 
connected by infinite 
Fibonacci lines running through the tiling. Both the density and the
Patterson data for different types of atomic positions are computed in closed form 
and shown to vary strongly between the planes.
\vspace{0.2cm}

In the planes we consider vertex positions of the tiling and
pentagonal positions arising from faces or top cuts through the Bergman
clusters. Changes of their correlation between the planes are
analyzed in three alternative models.  
The pentagonal holes observed in \cite{BU} admit
an interpretation in terms of these pentagons. 
If they arise from the Bergman top
pentagons, their observed larger size must reflect  a 
local reconstruction of the surface. 
If they arise from the Bergman top cut pentagons, 
they agree with  the observed size.

The geometry, the spacing, the density
and the  Patterson functions 
are computed from the bulk model. The sequence and spacing of terraces 
and the available Patterson data from \cite{BU} are well
reproduced. The observed Fibonacci string of planes yields 
information on the density of atomic positions. The observed size of the terraces shows some correlation  with 
the model structure in the planes. 

More detailed experimental studies of the terrace structure in
i-quasicrystals  are suggested.

\section{Acknowledgement.}
The authors would like to thank Tobias Kramer for substantial 
help in the computation and  presentation of data.

\end{document}

%% file: quadrat.tex
\begin{picture}(0,0)%
\special{psfile=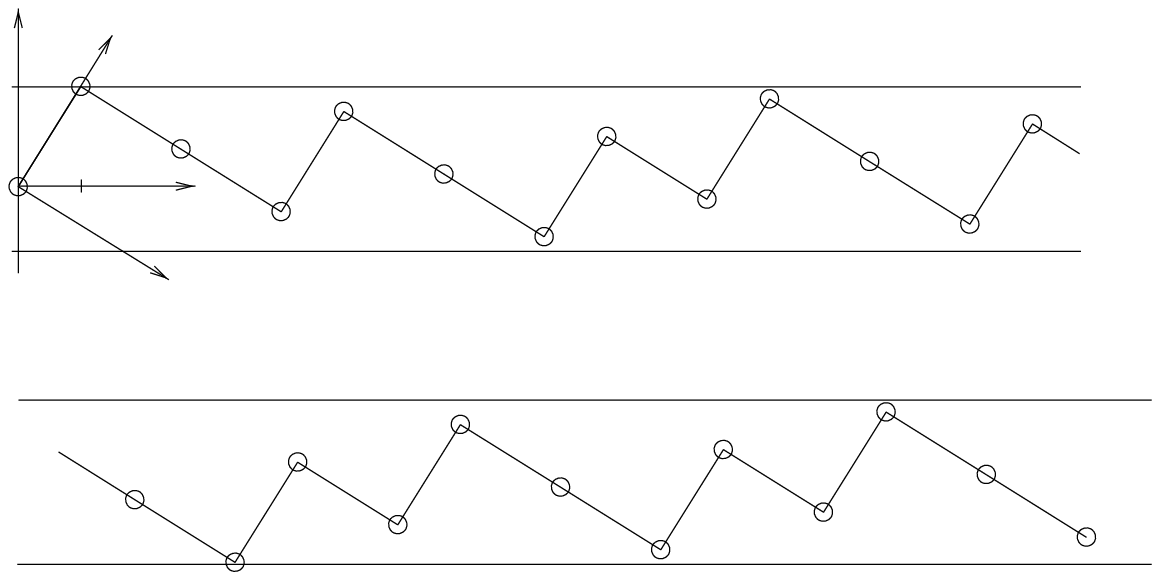}%
\end{picture}%
\setlength{\unitlength}{2486sp}%
\begingroup\makeatletter\ifx\SetFigFont\undefined
\def\x#1#2#3#4#5#6#7\relax{\def\x{#1#2#3#4#5#6}}%
\expandafter\x\fmtname xxxxxx\relax \def\y{splain}%
\ifx\x\y   
\gdef\SetFigFont#1#2#3{%
  \ifnum #1<17\tiny\else \ifnum #1<20\small\else
  \ifnum #1<24\normalsize\else \ifnum #1<29\large\else
  \ifnum #1<34\Large\else \ifnum #1<41\LARGE\else
     \huge\fi\fi\fi\fi\fi\fi
  \csname #3\endcsname}%
\else
\gdef\SetFigFont#1#2#3{\begingroup
  \count@#1\relax \ifnum 25<\count@\count@25\fi
  \def\x{\endgroup\@setsize\SetFigFont{#2pt}}%
  \expandafter\x
    \csname \romannumeral\the\count@ pt\expandafter\endcsname
    \csname @\romannumeral\the\count@ pt\endcsname
  \csname #3\endcsname}%
\fi
\fi\endgroup
\begin{picture}(8954,4968)(1003,-13678)
\put(1351,-11356){\makebox(0,0)[lb]{\smash{\SetFigFont{12}{14.4}{bf}0}}}
\put(1801,-11356){\makebox(0,0)[lb]{\smash{\SetFigFont{12}{14.4}{bf}1}}}
\put(2566,-11356){\makebox(0,0)[lb]{\smash{\SetFigFont{12}{14.4}{bf}2}}}
\put(3331,-11356){\makebox(0,0)[lb]{\smash{\SetFigFont{12}{14.4}{bf}3}}}
\put(3736,-11356){\makebox(0,0)[lb]{\smash{\SetFigFont{12}{14.4}{bf}4}}}
\put(4501,-11356){\makebox(0,0)[lb]{\smash{\SetFigFont{12}{14.4}{bf}5}}}
\put(5311,-11356){\makebox(0,0)[lb]{\smash{\SetFigFont{12}{14.4}{bf}6}}}
\put(6571,-11356){\makebox(0,0)[lb]{\smash{\SetFigFont{12}{14.4}{bf}8}}}
\put(7021,-11356){\makebox(0,0)[lb]{\smash{\SetFigFont{12}{14.4}{bf}9}}}
\put(7786,-11356){\makebox(0,0)[lb]{\smash{\SetFigFont{12}{14.4}{bf}10}}}
\put(8506,-11356){\makebox(0,0)[lb]{\smash{\SetFigFont{12}{14.4}{bf}11}}}
\put(9001,-11356){\makebox(0,0)[lb]{\smash{\SetFigFont{12}{14.4}{bf}12}}}
\put(2116,-13651){\makebox(0,0)[lb]{\smash{\SetFigFont{12}{14.4}{bf}13}}}
\put(3376,-13651){\makebox(0,0)[lb]{\smash{\SetFigFont{12}{14.4}{bf}15}}}
\put(4141,-13651){\makebox(0,0)[lb]{\smash{\SetFigFont{12}{14.4}{bf}16}}}
\put(4636,-13651){\makebox(0,0)[lb]{\smash{\SetFigFont{12}{14.4}{bf}17}}}
\put(5356,-13651){\makebox(0,0)[lb]{\smash{\SetFigFont{12}{14.4}{bf}18}}}
\put(6166,-13651){\makebox(0,0)[lb]{\smash{\SetFigFont{12}{14.4}{bf}19}}}
\put(7381,-13651){\makebox(0,0)[lb]{\smash{\SetFigFont{12}{14.4}{bf}21}}}
\put(8641,-13651){\makebox(0,0)[lb]{\smash{\SetFigFont{12}{14.4}{bf}23}}}
\put(9361,-13651){\makebox(0,0)[lb]{\smash{\SetFigFont{12}{14.4}{bf}24}}}
\put(5761,-11356){\makebox(0,0)[lb]{\smash{\SetFigFont{12}{14.4}{bf}7}}}
\put(2881,-13651){\makebox(0,0)[lb]{\smash{\SetFigFont{12}{14.4}{bf}14}}}
\put(6571,-13651){\makebox(0,0)[lb]{\smash{\SetFigFont{12}{14.4}{bf}20}}}
\put(7831,-13651){\makebox(0,0)[lb]{\smash{\SetFigFont{12}{14.4}{bf}22}}}
\put(2071,-9076){\makebox(0,0)[lb]{\smash{\SetFigFont{12}{14.4}{bf}$x_2$}}}
\put(1096,-8866){\makebox(0,0)[lb]{\smash{\SetFigFont{12}{14.4}{bf}$x_{\perp}$}}}
\put(2461,-10548){\makebox(0,0)[lb]{\smash{\SetFigFont{12}{14.4}{bf}$x_{\parallel}$}}}
\put(2519,-11047){\makebox(0,0)[lb]{\smash{\SetFigFont{12}{14.4}{bf}$x_1$}}}
\put(1003,-10890){\makebox(0,0)[lb]{\smash{\SetFigFont{9}{10.8}{bf}$-1$}}}
\put(1081,-9618){\makebox(0,0)[lb]{\smash{\SetFigFont{9}{10.8}{bf}$\tau$}}}
\put(1755,-10237){\makebox(0,0)[lb]{\smash{\SetFigFont{9}{10.8}{bf}$1$}}}
\end{picture}

%% file: w3.tex
\begin{picture}(0,0)%
\special{psfile=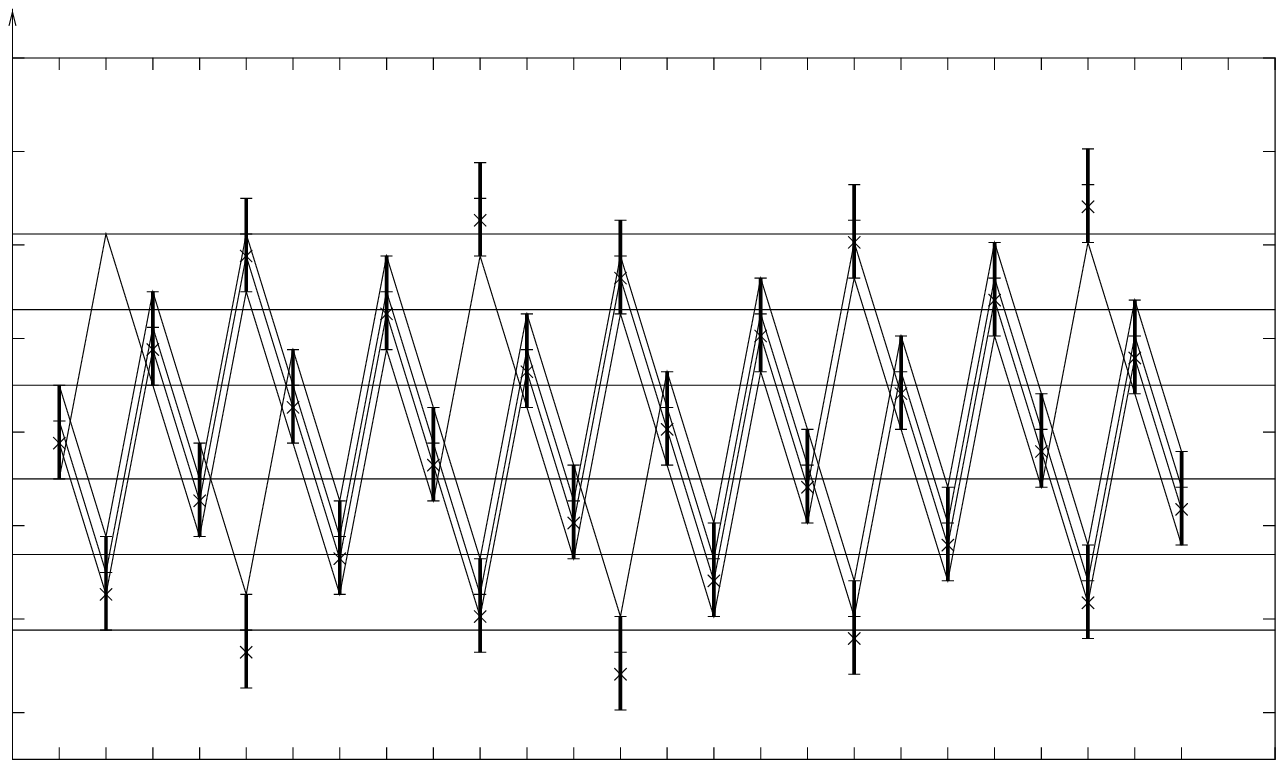}%
\end{picture}%
\setlength{\unitlength}{2171sp}%
\begingroup\makeatletter\ifx\SetFigFont\undefined
\def\x#1#2#3#4#5#6#7\relax{\def\x{#1#2#3#4#5#6}}%
\expandafter\x\fmtname xxxxxx\relax \def\y{splain}%
\ifx\x\y   
\gdef\SetFigFont#1#2#3{%
  \ifnum #1<17\tiny\else \ifnum #1<20\small\else
  \ifnum #1<24\normalsize\else \ifnum #1<29\large\else
  \ifnum #1<34\Large\else \ifnum #1<41\LARGE\else
     \huge\fi\fi\fi\fi\fi\fi
  \csname #3\endcsname}%
\else
\gdef\SetFigFont#1#2#3{\begingroup
  \count@#1\relax \ifnum 25<\count@\count@25\fi
  \def\x{\endgroup\@setsize\SetFigFont{#2pt}}%
  \expandafter\x
    \csname \romannumeral\the\count@ pt\expandafter\endcsname
    \csname @\romannumeral\the\count@ pt\endcsname
  \csname #3\endcsname}%
\fi
\fi\endgroup
\begin{picture}(11625,7052)(376,-6267)
\put(570,109){\makebox(0,0)[lb]{\smash{\SetFigFont{12}{14.4}{bf}$4$}}}
\put(570,-708){\makebox(0,0)[lb]{\smash{\SetFigFont{12}{14.4}{bf}$3$}}}
\put(570,-1523){\makebox(0,0)[lb]{\smash{\SetFigFont{12}{14.4}{bf}$2$}}}
\put(570,-2340){\makebox(0,0)[lb]{\smash{\SetFigFont{12}{14.4}{bf}$1$}}}
\put(570,-3156){\makebox(0,0)[lb]{\smash{\SetFigFont{12}{14.4}{bf}$0$}}}
\put(376,-5611){\makebox(0,0)[lb]{\smash{\SetFigFont{12}{14.4}{bf}$-3$}}}
\put(376,-4786){\makebox(0,0)[lb]{\smash{\SetFigFont{12}{14.4}{bf}$-2$}}}
\put(376,-3961){\makebox(0,0)[lb]{\smash{\SetFigFont{12}{14.4}{bf}$-1$}}}
\put(12001,-1411){\makebox(0,0)[lb]{\smash{\SetFigFont{12}{14.4}{bf}$\frac{1}{2}\tau^3$}}}
\put(12001,-2086){\makebox(0,0)[lb]{\smash{\SetFigFont{12}{14.4}{bf}$\frac{1}{2}\tau^2$}}}
\put(12001,-2761){\makebox(0,0)[lb]{\smash{\SetFigFont{12}{14.4}{bf}$\frac{1}{2}$}}}
\put(12001,-3586){\makebox(0,0)[lb]{\smash{\SetFigFont{12}{14.4}{bf}$-\frac{1}{2}$}}}
\put(12001,-4186){\makebox(0,0)[lb]{\smash{\SetFigFont{12}{14.4}{bf}$-\frac{1}{2}\tau^2$}}}
\put(12001,-4861){\makebox(0,0)[lb]{\smash{\SetFigFont{12}{14.4}{bf}$-\frac{1}{2}\tau^3$}}}
\put(1183,-6246){\makebox(0,0)[lb]{\smash{\SetFigFont{10}{12.0}{bf}0}}}
\put(1592,-6246){\makebox(0,0)[lb]{\smash{\SetFigFont{10}{12.0}{bf}1}}}
\put(2000,-6246){\makebox(0,0)[lb]{\smash{\SetFigFont{10}{12.0}{bf}2}}}
\put(2408,-6246){\makebox(0,0)[lb]{\smash{\SetFigFont{10}{12.0}{bf}3}}}
\put(2815,-6246){\makebox(0,0)[lb]{\smash{\SetFigFont{10}{12.0}{bf}4}}}
\put(3223,-6246){\makebox(0,0)[lb]{\smash{\SetFigFont{10}{12.0}{bf}5}}}
\put(3632,-6246){\makebox(0,0)[lb]{\smash{\SetFigFont{10}{12.0}{bf}6}}}
\put(4040,-6246){\makebox(0,0)[lb]{\smash{\SetFigFont{10}{12.0}{bf}7}}}
\put(4448,-6246){\makebox(0,0)[lb]{\smash{\SetFigFont{10}{12.0}{bf}8}}}
\put(4857,-6246){\makebox(0,0)[lb]{\smash{\SetFigFont{10}{12.0}{bf}9}}}
\put(5200,-6246){\makebox(0,0)[lb]{\smash{\SetFigFont{10}{12.0}{bf}10}}}
\put(5608,-6246){\makebox(0,0)[lb]{\smash{\SetFigFont{10}{12.0}{bf}11}}}
\put(6017,-6246){\makebox(0,0)[lb]{\smash{\SetFigFont{10}{12.0}{bf}12}}}
\put(6423,-6246){\makebox(0,0)[lb]{\smash{\SetFigFont{10}{12.0}{bf}13}}}
\put(6832,-6246){\makebox(0,0)[lb]{\smash{\SetFigFont{10}{12.0}{bf}14}}}
\put(7240,-6246){\makebox(0,0)[lb]{\smash{\SetFigFont{10}{12.0}{bf}15}}}
\put(7648,-6246){\makebox(0,0)[lb]{\smash{\SetFigFont{10}{12.0}{bf}16}}}
\put(8057,-6246){\makebox(0,0)[lb]{\smash{\SetFigFont{10}{12.0}{bf}17}}}
\put(8465,-6246){\makebox(0,0)[lb]{\smash{\SetFigFont{10}{12.0}{bf}18}}}
\put(9282,-6246){\makebox(0,0)[lb]{\smash{\SetFigFont{10}{12.0}{bf}20}}}
\put(8873,-6246){\makebox(0,0)[lb]{\smash{\SetFigFont{10}{12.0}{bf}19}}}
\put(9690,-6246){\makebox(0,0)[lb]{\smash{\SetFigFont{10}{12.0}{bf}21}}}
\put(10097,-6246){\makebox(0,0)[lb]{\smash{\SetFigFont{10}{12.0}{bf}22}}}
\put(10505,-6246){\makebox(0,0)[lb]{\smash{\SetFigFont{10}{12.0}{bf}23}}}
\put(10913,-6246){\makebox(0,0)[lb]{\smash{\SetFigFont{10}{12.0}{bf}24}}}
\put(980,629){\makebox(0,0)[lb]{\smash{\SetFigFont{12}{14.4}{bf}$y_{\perp}$}}}
\end{picture}

%% file: axes.tex
\begin{picture}(0,0)%
\includegraphics{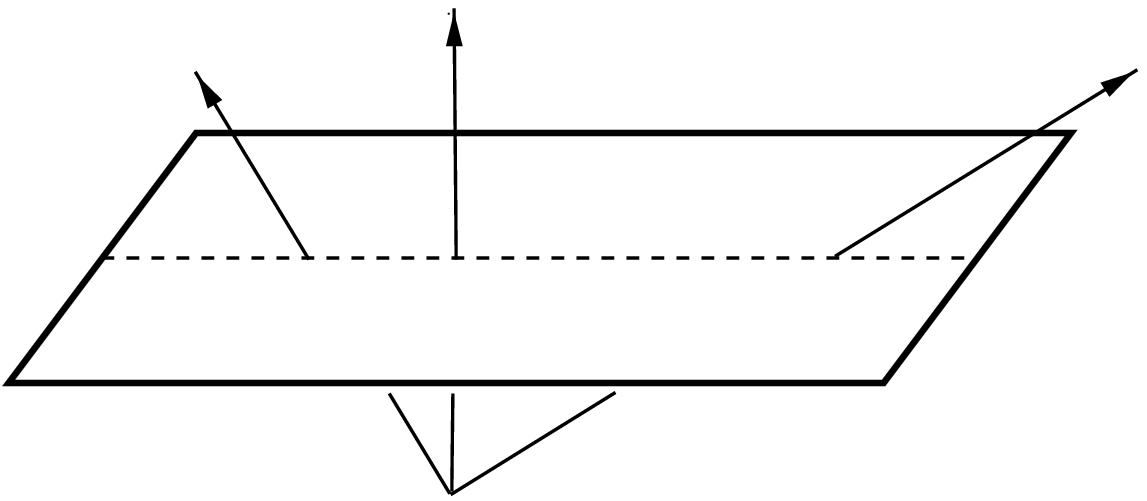}%
\end{picture}%
\setlength{\unitlength}{3947sp}%
\begingroup\makeatletter\ifx\SetFigFont\undefined
\def\x#1#2#3#4#5#6#7\relax{\def\x{#1#2#3#4#5#6}}%
\expandafter\x\fmtname xxxxxx\relax \def\y{splain}%
\ifx\x\y   
\gdef\SetFigFont#1#2#3{%
  \ifnum #1<17\tiny\else \ifnum #1<20\small\else
  \ifnum #1<24\normalsize\else \ifnum #1<29\large\else
  \ifnum #1<34\Large\else \ifnum #1<41\LARGE\else
     \huge\fi\fi\fi\fi\fi\fi
  \csname #3\endcsname}%
\else
\gdef\SetFigFont#1#2#3{\begingroup
  \count@#1\relax \ifnum 25<\count@\count@25\fi
  \def\x{\endgroup\@setsize\SetFigFont{#2pt}}%
  \expandafter\x
    \csname \romannumeral\the\count@ pt\expandafter\endcsname
    \csname @\romannumeral\the\count@ pt\endcsname
  \csname #3\endcsname}%
\fi
\fi\endgroup
\begin{picture}(5472,2415)(2743,-3019)
\put(3224,-1031){\makebox(0,0)[lb]{\smash{\SetFigFont{12}{14.4}{bf}$2'$}}}
\put(5200,-736){\makebox(0,0)[lb]{\smash{\SetFigFont{12}{14.4}{bf}$5$}}}
\put(7473,-1015){\makebox(0,0)[lb]{\smash{\SetFigFont{12}{14.4}{bf}$2$}}}
\end{picture}

%% file: guertel.tex
\begin{picture}(0,0)%
\includegraphics{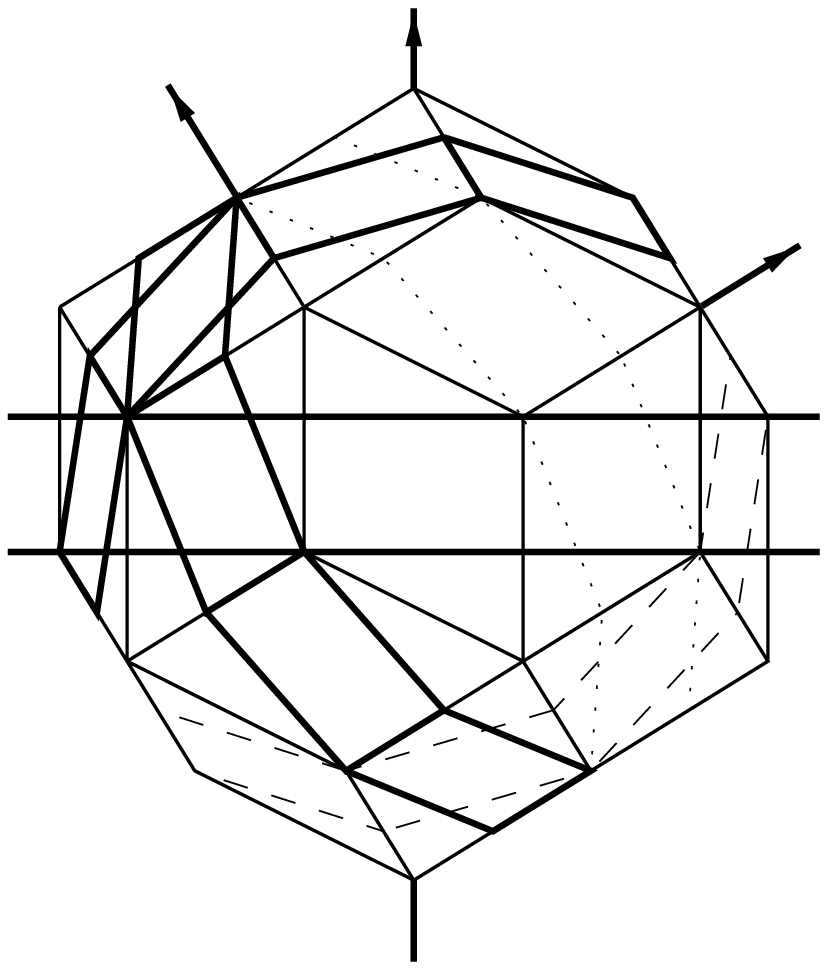}%
\end{picture}%
\setlength{\unitlength}{3947sp}%
\begingroup\makeatletter\ifx\SetFigFont\undefined
\def\x#1#2#3#4#5#6#7\relax{\def\x{#1#2#3#4#5#6}}%
\expandafter\x\fmtname xxxxxx\relax \def\y{splain}%
\ifx\x\y   
\gdef\SetFigFont#1#2#3{%
  \ifnum #1<17\tiny\else \ifnum #1<20\small\else
  \ifnum #1<24\normalsize\else \ifnum #1<29\large\else
  \ifnum #1<34\Large\else \ifnum #1<41\LARGE\else
     \huge\fi\fi\fi\fi\fi\fi
  \csname #3\endcsname}%
\else
\gdef\SetFigFont#1#2#3{\begingroup
  \count@#1\relax \ifnum 25<\count@\count@25\fi
  \def\x{\endgroup\@setsize\SetFigFont{#2pt}}%
  \expandafter\x
    \csname \romannumeral\the\count@ pt\expandafter\endcsname
    \csname @\romannumeral\the\count@ pt\endcsname
  \csname #3\endcsname}%
\fi
\fi\endgroup
\begin{picture}(4117,4717)(2906,-5316)
\put(3408,-1174){\makebox(0,0)[lb]{\smash{\SetFigFont{12}{14.4}{bf}$2$}}}
\put(5208,-731){\makebox(0,0)[lb]{\smash{\SetFigFont{12}{14.4}{bf}$5$}}}
\put(7023,-2081){\makebox(0,0)[lb]{\smash{\SetFigFont{12}{14.4}{bf}$2'$}}}
\end{picture}

%% file: w42.tex
\begin{picture}(0,0)%
\special{psfile=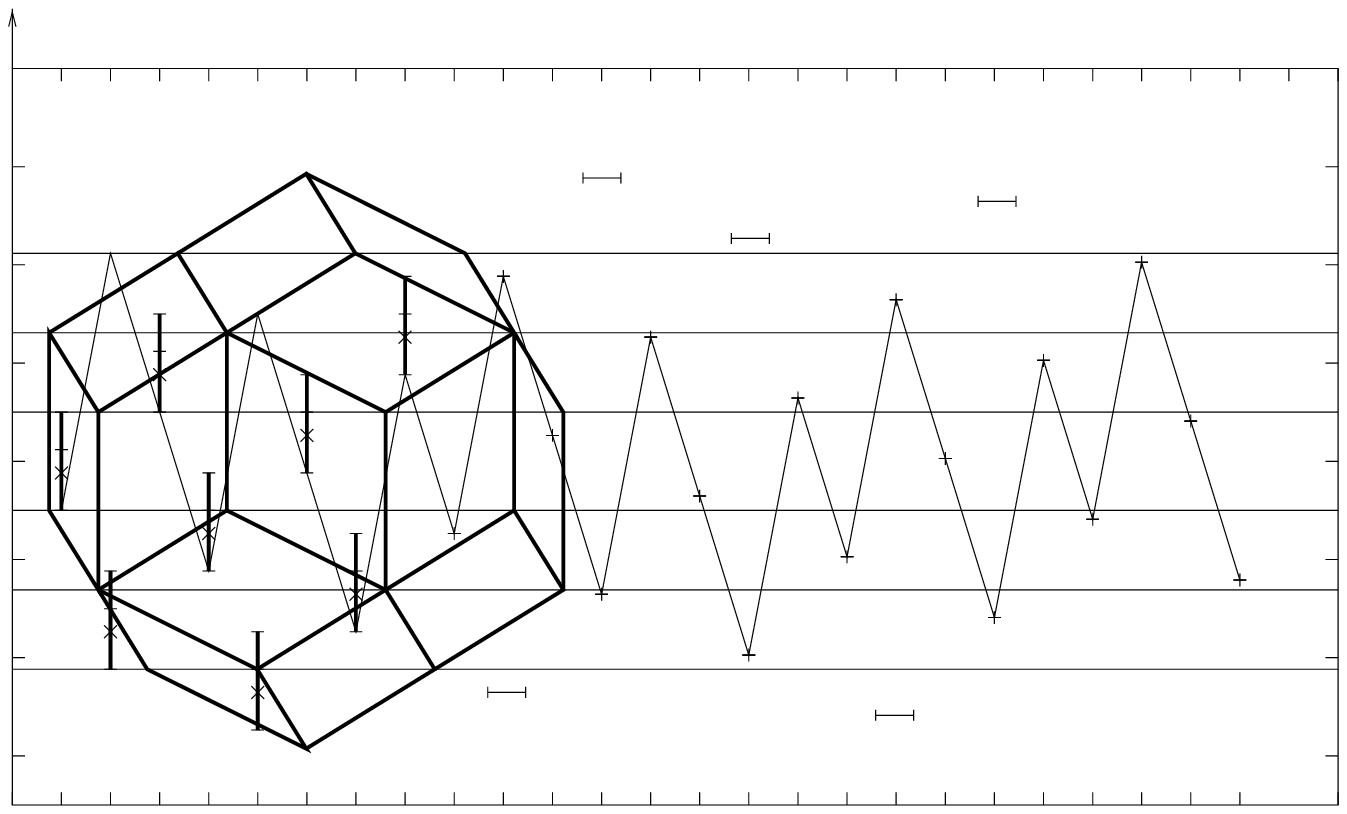}%
\end{picture}%
\setlength{\unitlength}{2279sp}%
\begingroup\makeatletter\ifx\SetFigFont\undefined
\def\x#1#2#3#4#5#6#7\relax{\def\x{#1#2#3#4#5#6}}%
\expandafter\x\fmtname xxxxxx\relax \def\y{splain}%
\ifx\x\y   
\gdef\SetFigFont#1#2#3{%
  \ifnum #1<17\tiny\else \ifnum #1<20\small\else
  \ifnum #1<24\normalsize\else \ifnum #1<29\large\else
  \ifnum #1<34\Large\else \ifnum #1<41\LARGE\else
     \huge\fi\fi\fi\fi\fi\fi
  \csname #3\endcsname}%
\else
\gdef\SetFigFont#1#2#3{\begingroup
  \count@#1\relax \ifnum 25<\count@\count@25\fi
  \def\x{\endgroup\@setsize\SetFigFont{#2pt}}%
  \expandafter\x
    \csname \romannumeral\the\count@ pt\expandafter\endcsname
    \csname @\romannumeral\the\count@ pt\endcsname
  \csname #3\endcsname}%
\fi
\fi\endgroup
\begin{picture}(11551,7436)(751,-6744)
\put(5917,-6396){\makebox(0,0)[lb]{\smash{\SetFigFont{10}{12.0}{bf}11}}}
\put(1901,-6396){\makebox(0,0)[lb]{\smash{\SetFigFont{10}{12.0}{bf}1}}}
\put(2309,-6396){\makebox(0,0)[lb]{\smash{\SetFigFont{10}{12.0}{bf}2}}}
\put(2717,-6396){\makebox(0,0)[lb]{\smash{\SetFigFont{10}{12.0}{bf}3}}}
\put(3124,-6396){\makebox(0,0)[lb]{\smash{\SetFigFont{10}{12.0}{bf}4}}}
\put(3532,-6396){\makebox(0,0)[lb]{\smash{\SetFigFont{10}{12.0}{bf}5}}}
\put(3941,-6396){\makebox(0,0)[lb]{\smash{\SetFigFont{10}{12.0}{bf}6}}}
\put(4349,-6396){\makebox(0,0)[lb]{\smash{\SetFigFont{10}{12.0}{bf}7}}}
\put(4757,-6396){\makebox(0,0)[lb]{\smash{\SetFigFont{10}{12.0}{bf}8}}}
\put(5166,-6396){\makebox(0,0)[lb]{\smash{\SetFigFont{10}{12.0}{bf}9}}}
\put(5509,-6396){\makebox(0,0)[lb]{\smash{\SetFigFont{10}{12.0}{bf}10}}}
\put(879,-41){\makebox(0,0)[lb]{\smash{\SetFigFont{12}{14.4}{bf}$4$}}}
\put(879,-858){\makebox(0,0)[lb]{\smash{\SetFigFont{12}{14.4}{bf}$3$}}}
\put(879,-1673){\makebox(0,0)[lb]{\smash{\SetFigFont{12}{14.4}{bf}$2$}}}
\put(879,-2490){\makebox(0,0)[lb]{\smash{\SetFigFont{12}{14.4}{bf}$1$}}}
\put(879,-3306){\makebox(0,0)[lb]{\smash{\SetFigFont{12}{14.4}{bf}$0$}}}
\put(751,-4111){\makebox(0,0)[lb]{\smash{\SetFigFont{12}{14.4}{bf}$-1$}}}
\put(751,-4936){\makebox(0,0)[lb]{\smash{\SetFigFont{12}{14.4}{bf}$-2$}}}
\put(751,-5761){\makebox(0,0)[lb]{\smash{\SetFigFont{12}{14.4}{bf}$-3$}}}
\put(1492,-6396){\makebox(0,0)[lb]{\smash{\SetFigFont{10}{12.0}{bf}0}}}
\put(6326,-6396){\makebox(0,0)[lb]{\smash{\SetFigFont{10}{12.0}{bf}12}}}
\put(6732,-6396){\makebox(0,0)[lb]{\smash{\SetFigFont{10}{12.0}{bf}13}}}
\put(7141,-6396){\makebox(0,0)[lb]{\smash{\SetFigFont{10}{12.0}{bf}14}}}
\put(7549,-6396){\makebox(0,0)[lb]{\smash{\SetFigFont{10}{12.0}{bf}15}}}
\put(7957,-6396){\makebox(0,0)[lb]{\smash{\SetFigFont{10}{12.0}{bf}16}}}
\put(8366,-6396){\makebox(0,0)[lb]{\smash{\SetFigFont{10}{12.0}{bf}17}}}
\put(8774,-6396){\makebox(0,0)[lb]{\smash{\SetFigFont{10}{12.0}{bf}18}}}
\put(9182,-6396){\makebox(0,0)[lb]{\smash{\SetFigFont{10}{12.0}{bf}19}}}
\put(9591,-6396){\makebox(0,0)[lb]{\smash{\SetFigFont{10}{12.0}{bf}20}}}
\put(9999,-6396){\makebox(0,0)[lb]{\smash{\SetFigFont{10}{12.0}{bf}21}}}
\put(10406,-6396){\makebox(0,0)[lb]{\smash{\SetFigFont{10}{12.0}{bf}22}}}
\put(10814,-6396){\makebox(0,0)[lb]{\smash{\SetFigFont{10}{12.0}{bf}23}}}
\put(11222,-6396){\makebox(0,0)[lb]{\smash{\SetFigFont{10}{12.0}{bf}24}}}
\put(12302,-1562){\makebox(0,0)[lb]{\smash{\SetFigFont{12}{14.4}{bf}$\frac{1}{2}\tau^3$}}}
\put(12302,-2237){\makebox(0,0)[lb]{\smash{\SetFigFont{12}{14.4}{bf}$\frac{1}{2}\tau^2$}}}
\put(12302,-2912){\makebox(0,0)[lb]{\smash{\SetFigFont{12}{14.4}{bf}$\frac{1}{2}$}}}
\put(12302,-3737){\makebox(0,0)[lb]{\smash{\SetFigFont{12}{14.4}{bf}$-\frac{1}{2}$}}}
\put(12302,-4337){\makebox(0,0)[lb]{\smash{\SetFigFont{12}{14.4}{bf}$-\frac{1}{2}\tau^2$}}}
\put(12302,-5012){\makebox(0,0)[lb]{\smash{\SetFigFont{12}{14.4}{bf}$-\frac{1}{2}\tau^3$}}}
\put(5992,-6723){\makebox(0,0)[lb]{\smash{\SetFigFont{10}{12.0}{it}3}}}
\put(6399,-6723){\makebox(0,0)[lb]{\smash{\SetFigFont{10}{12.0}{it}4}}}
\put(6807,-6723){\makebox(0,0)[lb]{\smash{\SetFigFont{10}{12.0}{it}5}}}
\put(7216,-6723){\makebox(0,0)[lb]{\smash{\SetFigFont{10}{12.0}{it}6}}}
\put(7624,-6723){\makebox(0,0)[lb]{\smash{\SetFigFont{10}{12.0}{it}7}}}
\put(8032,-6723){\makebox(0,0)[lb]{\smash{\SetFigFont{10}{12.0}{it}8}}}
\put(8441,-6723){\makebox(0,0)[lb]{\smash{\SetFigFont{10}{12.0}{it}9}}}
\put(8784,-6723){\makebox(0,0)[lb]{\smash{\SetFigFont{10}{12.0}{it}10}}}
\put(9192,-6723){\makebox(0,0)[lb]{\smash{\SetFigFont{10}{12.0}{it}11}}}
\put(5584,-6723){\makebox(0,0)[lb]{\smash{\SetFigFont{10}{12.0}{it}2}}}
\put(5176,-6723){\makebox(0,0)[lb]{\smash{\SetFigFont{10}{12.0}{it}1}}}
\put(1292,536){\makebox(0,0)[lb]{\smash{\SetFigFont{12}{14.4}{bf}$y_{\perp}$}}}
\put(5971,-788){\makebox(0,0)[lb]{\smash{\SetFigFont{12}{14.4}{it}$3^+$}}}
\put(5131,-5416){\makebox(0,0)[lb]{\smash{\SetFigFont{12}{14.4}{it}$1^-$}}}
\put(8371,-5641){\makebox(0,0)[lb]{\smash{\SetFigFont{12}{14.4}{it}$9^-$}}}
\put(7205,-1279){\makebox(0,0)[lb]{\smash{\SetFigFont{12}{14.4}{it}$6^+$}}}
\put(9223,-975){\makebox(0,0)[lb]{\smash{\SetFigFont{12}{14.4}{it}$11^+$}}}
\end{picture}

%% file: dodek.tex
\begin{picture}(0,0)%
\includegraphics{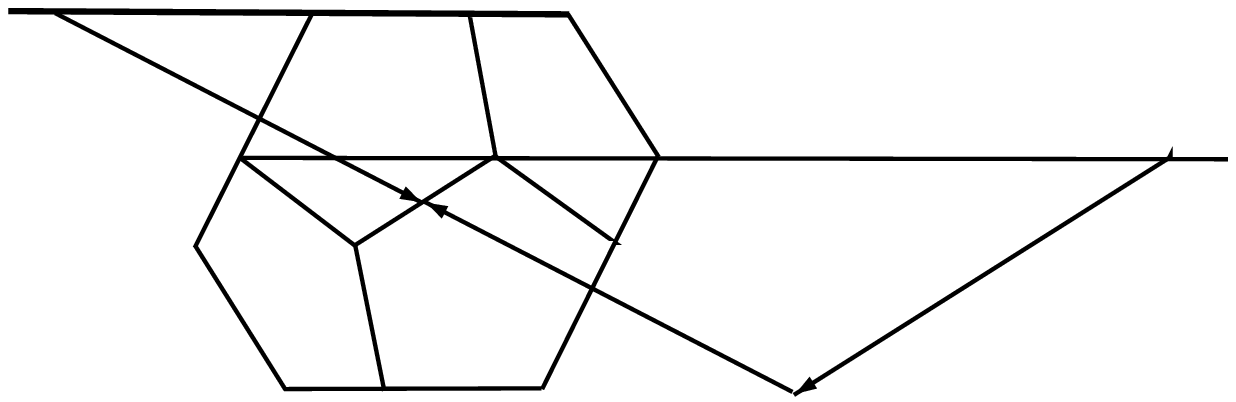}%
\end{picture}%
\setlength{\unitlength}{2565sp}%
\begingroup\makeatletter\ifx\SetFigFont\undefined
\def\x#1#2#3#4#5#6#7\relax{\def\x{#1#2#3#4#5#6}}%
\expandafter\x\fmtname xxxxxx\relax \def\y{splain}%
\ifx\x\y   
\gdef\SetFigFont#1#2#3{%
  \ifnum #1<17\tiny\else \ifnum #1<20\small\else
  \ifnum #1<24\normalsize\else \ifnum #1<29\large\else
  \ifnum #1<34\Large\else \ifnum #1<41\LARGE\else
     \huge\fi\fi\fi\fi\fi\fi
  \csname #3\endcsname}%
\else
\gdef\SetFigFont#1#2#3{\begingroup
  \count@#1\relax \ifnum 25<\count@\count@25\fi
  \def\x{\endgroup\@setsize\SetFigFont{#2pt}}%
  \expandafter\x
    \csname \romannumeral\the\count@ pt\expandafter\endcsname
    \csname @\romannumeral\the\count@ pt\endcsname
  \csname #3\endcsname}%
\fi
\fi\endgroup
\begin{picture}(9122,2911)(1726,-4449)
\put(1726,-2461){\makebox(0,0)[lb]{\smash{\SetFigFont{12}{14.4}{bf}$e_{5\parallel}$}}}
\put(9151,-4036){\makebox(0,0)[lb]{\smash{\SetFigFont{12}{14.4}{bf}$(e_1-e_5)_{\parallel}$}}}
\end{picture}

%% file: treppe1.tex
\begin{picture}(0,0)%
\special{psfile=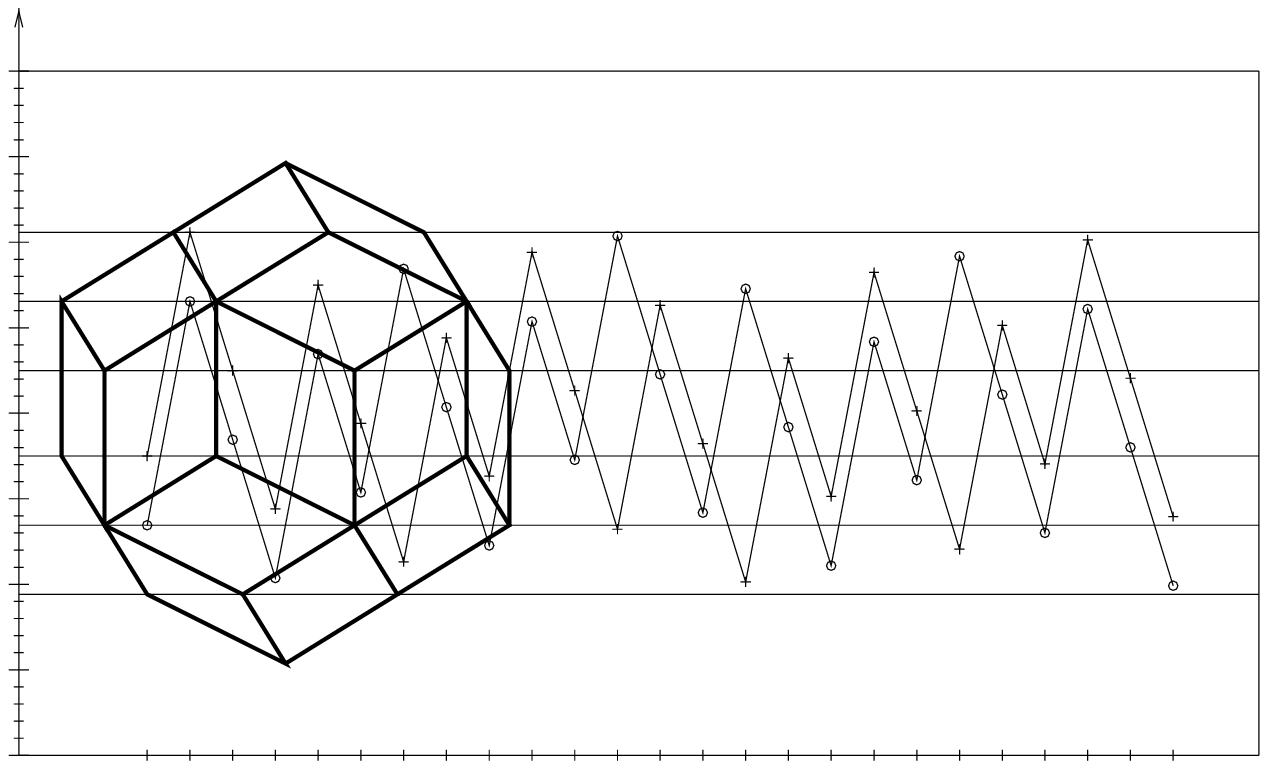}%
\end{picture}%
\setlength{\unitlength}{2447sp}%
\begingroup\makeatletter\ifx\SetFigFont\undefined
\def\x#1#2#3#4#5#6#7\relax{\def\x{#1#2#3#4#5#6}}%
\expandafter\x\fmtname xxxxxx\relax \def\y{splain}%
\ifx\x\y   
\gdef\SetFigFont#1#2#3{%
  \ifnum #1<17\tiny\else \ifnum #1<20\small\else
  \ifnum #1<24\normalsize\else \ifnum #1<29\large\else
  \ifnum #1<34\Large\else \ifnum #1<41\LARGE\else
     \huge\fi\fi\fi\fi\fi\fi
  \csname #3\endcsname}%
\else
\gdef\SetFigFont#1#2#3{\begingroup
  \count@#1\relax \ifnum 25<\count@\count@25\fi
  \def\x{\endgroup\@setsize\SetFigFont{#2pt}}%
  \expandafter\x
    \csname \romannumeral\the\count@ pt\expandafter\endcsname
    \csname @\romannumeral\the\count@ pt\endcsname
  \csname #3\endcsname}%
\fi
\fi\endgroup
\begin{picture}(10060,6286)(1126,-6826)
\put(1276,-1261){\makebox(0,0)[lb]{\smash{\SetFigFont{12}{14.4}{bf}$4$}}}
\put(1276,-1936){\makebox(0,0)[lb]{\smash{\SetFigFont{12}{14.4}{bf}$3$}}}
\put(1276,-2611){\makebox(0,0)[lb]{\smash{\SetFigFont{12}{14.4}{bf}$2$}}}
\put(1276,-3286){\makebox(0,0)[lb]{\smash{\SetFigFont{12}{14.4}{bf}$1$}}}
\put(1276,-3961){\makebox(0,0)[lb]{\smash{\SetFigFont{12}{14.4}{bf}$0$}}}
\put(1126,-6586){\makebox(0,0)[lb]{\smash{\SetFigFont{12}{14.4}{bf}$-4$}}}
\put(1126,-5911){\makebox(0,0)[lb]{\smash{\SetFigFont{12}{14.4}{bf}$-3$}}}
\put(1126,-5236){\makebox(0,0)[lb]{\smash{\SetFigFont{12}{14.4}{bf}$-2$}}}
\put(1126,-4636){\makebox(0,0)[lb]{\smash{\SetFigFont{12}{14.4}{bf}$-1$}}}
\put(1652,-696){\makebox(0,0)[lb]{\smash{\SetFigFont{12}{14.4}{bf}$y_{\perp}$}}}
\put(4501,-6811){\makebox(0,0)[lb]{\smash{\SetFigFont{8}{9.6}{bf}6}}}
\put(4801,-6811){\makebox(0,0)[lb]{\smash{\SetFigFont{8}{9.6}{bf}7}}}
\put(4126,-6811){\makebox(0,0)[lb]{\smash{\SetFigFont{8}{9.6}{bf}5}}}
\put(3826,-6811){\makebox(0,0)[lb]{\smash{\SetFigFont{8}{9.6}{bf}4}}}
\put(3526,-6811){\makebox(0,0)[lb]{\smash{\SetFigFont{8}{9.6}{bf}3}}}
\put(3151,-6811){\makebox(0,0)[lb]{\smash{\SetFigFont{8}{9.6}{bf}2}}}
\put(2851,-6811){\makebox(0,0)[lb]{\smash{\SetFigFont{8}{9.6}{bf}1}}}
\put(5176,-6811){\makebox(0,0)[lb]{\smash{\SetFigFont{8}{9.6}{bf}8}}}
\put(10501,-6811){\makebox(0,0)[lb]{\smash{\SetFigFont{8}{9.6}{bf}24}}}
\put(10148,-6811){\makebox(0,0)[lb]{\smash{\SetFigFont{8}{9.6}{bf}23}}}
\put(9804,-6811){\makebox(0,0)[lb]{\smash{\SetFigFont{8}{9.6}{bf}22}}}
\put(9496,-6811){\makebox(0,0)[lb]{\smash{\SetFigFont{8}{9.6}{bf}21}}}
\put(9166,-6811){\makebox(0,0)[lb]{\smash{\SetFigFont{8}{9.6}{bf}20}}}
\put(8791,-6811){\makebox(0,0)[lb]{\smash{\SetFigFont{8}{9.6}{bf}19}}}
\put(8454,-6811){\makebox(0,0)[lb]{\smash{\SetFigFont{8}{9.6}{bf}18}}}
\put(8123,-6811){\makebox(0,0)[lb]{\smash{\SetFigFont{8}{9.6}{bf}17}}}
\put(7809,-6811){\makebox(0,0)[lb]{\smash{\SetFigFont{8}{9.6}{bf}16}}}
\put(7479,-6811){\makebox(0,0)[lb]{\smash{\SetFigFont{8}{9.6}{bf}15}}}
\put(7140,-6811){\makebox(0,0)[lb]{\smash{\SetFigFont{8}{9.6}{bf}14}}}
\put(6811,-6811){\makebox(0,0)[lb]{\smash{\SetFigFont{8}{9.6}{bf}13}}}
\put(6481,-6811){\makebox(0,0)[lb]{\smash{\SetFigFont{8}{9.6}{bf}12}}}
\put(6166,-6811){\makebox(0,0)[lb]{\smash{\SetFigFont{8}{9.6}{bf}11}}}
\put(5806,-6811){\makebox(0,0)[lb]{\smash{\SetFigFont{8}{9.6}{bf}10}}}
\put(5521,-6811){\makebox(0,0)[lb]{\smash{\SetFigFont{8}{9.6}{bf}9}}}
\put(2551,-6811){\makebox(0,0)[lb]{\smash{\SetFigFont{8}{9.6}{bf}0}}}
\end{picture}

%% file: treppe2.tex
\begin{picture}(0,0)%
\special{psfile=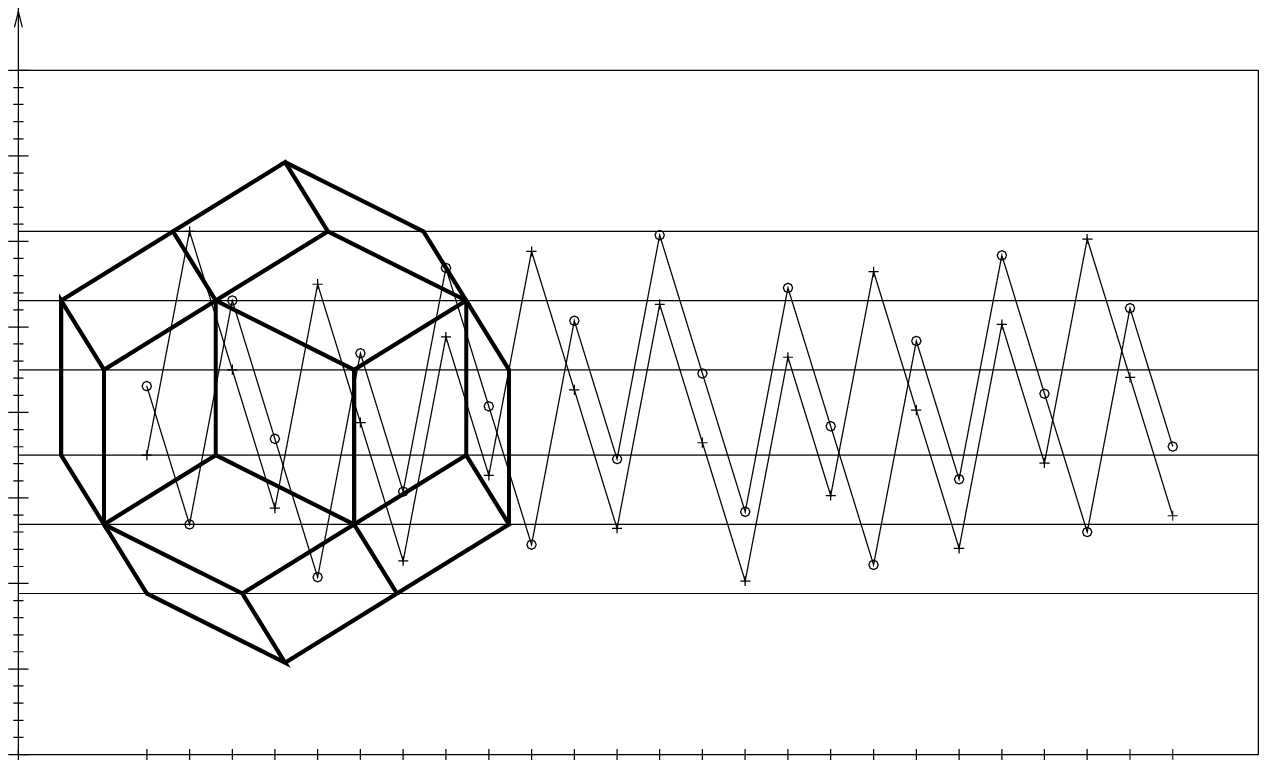}%
\end{picture}%
\setlength{\unitlength}{2447sp}%
\begingroup\makeatletter\ifx\SetFigFont\undefined
\def\x#1#2#3#4#5#6#7\relax{\def\x{#1#2#3#4#5#6}}%
\expandafter\x\fmtname xxxxxx\relax \def\y{splain}%
\ifx\x\y   
\gdef\SetFigFont#1#2#3{%
  \ifnum #1<17\tiny\else \ifnum #1<20\small\else
  \ifnum #1<24\normalsize\else \ifnum #1<29\large\else
  \ifnum #1<34\Large\else \ifnum #1<41\LARGE\else
     \huge\fi\fi\fi\fi\fi\fi
  \csname #3\endcsname}%
\else
\gdef\SetFigFont#1#2#3{\begingroup
  \count@#1\relax \ifnum 25<\count@\count@25\fi
  \def\x{\endgroup\@setsize\SetFigFont{#2pt}}%
  \expandafter\x
    \csname \romannumeral\the\count@ pt\expandafter\endcsname
    \csname @\romannumeral\the\count@ pt\endcsname
  \csname #3\endcsname}%
\fi
\fi\endgroup
\begin{picture}(10134,6171)(1126,-6976)
\put(1276,-2761){\makebox(0,0)[lb]{\smash{\SetFigFont{12}{14.4}{bf}$2$}}}
\put(1276,-2086){\makebox(0,0)[lb]{\smash{\SetFigFont{12}{14.4}{bf}$3$}}}
\put(1276,-1411){\makebox(0,0)[lb]{\smash{\SetFigFont{12}{14.4}{bf}$4$}}}
\put(1276,-3436){\makebox(0,0)[lb]{\smash{\SetFigFont{12}{14.4}{bf}$1$}}}
\put(1276,-4111){\makebox(0,0)[lb]{\smash{\SetFigFont{12}{14.4}{bf}$0$}}}
\put(1126,-4711){\makebox(0,0)[lb]{\smash{\SetFigFont{12}{14.4}{bf}$-1$}}}
\put(1126,-5386){\makebox(0,0)[lb]{\smash{\SetFigFont{12}{14.4}{bf}$-2$}}}
\put(1126,-6061){\makebox(0,0)[lb]{\smash{\SetFigFont{12}{14.4}{bf}$-3$}}}
\put(1126,-6736){\makebox(0,0)[lb]{\smash{\SetFigFont{12}{14.4}{bf}$-4$}}}
\put(1801,-961){\makebox(0,0)[lb]{\smash{\SetFigFont{12}{14.4}{bf}$y_{\perp}$}}}
\put(3226,-6961){\makebox(0,0)[lb]{\smash{\SetFigFont{8}{9.6}{bf}2}}}
\put(3601,-6961){\makebox(0,0)[lb]{\smash{\SetFigFont{8}{9.6}{bf}3}}}
\put(3901,-6961){\makebox(0,0)[lb]{\smash{\SetFigFont{8}{9.6}{bf}4}}}
\put(4276,-6961){\makebox(0,0)[lb]{\smash{\SetFigFont{8}{9.6}{bf}5}}}
\put(4576,-6961){\makebox(0,0)[lb]{\smash{\SetFigFont{8}{9.6}{bf}6}}}
\put(4951,-6961){\makebox(0,0)[lb]{\smash{\SetFigFont{8}{9.6}{bf}7}}}
\put(5251,-6961){\makebox(0,0)[lb]{\smash{\SetFigFont{8}{9.6}{bf}8}}}
\put(5551,-6961){\makebox(0,0)[lb]{\smash{\SetFigFont{8}{9.6}{bf}9}}}
\put(5926,-6961){\makebox(0,0)[lb]{\smash{\SetFigFont{8}{9.6}{bf}10}}}
\put(6226,-6961){\makebox(0,0)[lb]{\smash{\SetFigFont{8}{9.6}{bf}11}}}
\put(7201,-6961){\makebox(0,0)[lb]{\smash{\SetFigFont{8}{9.6}{bf}14}}}
\put(8551,-6961){\makebox(0,0)[lb]{\smash{\SetFigFont{8}{9.6}{bf}18}}}
\put(9526,-6961){\makebox(0,0)[lb]{\smash{\SetFigFont{8}{9.6}{bf}21}}}
\put(6548,-6961){\makebox(0,0)[lb]{\smash{\SetFigFont{8}{9.6}{bf}12}}}
\put(6886,-6961){\makebox(0,0)[lb]{\smash{\SetFigFont{8}{9.6}{bf}13}}}
\put(7554,-6961){\makebox(0,0)[lb]{\smash{\SetFigFont{8}{9.6}{bf}15}}}
\put(7883,-6961){\makebox(0,0)[lb]{\smash{\SetFigFont{8}{9.6}{bf}16}}}
\put(8214,-6961){\makebox(0,0)[lb]{\smash{\SetFigFont{8}{9.6}{bf}17}}}
\put(8866,-6961){\makebox(0,0)[lb]{\smash{\SetFigFont{8}{9.6}{bf}19}}}
\put(9189,-6961){\makebox(0,0)[lb]{\smash{\SetFigFont{8}{9.6}{bf}20}}}
\put(9871,-6961){\makebox(0,0)[lb]{\smash{\SetFigFont{8}{9.6}{bf}22}}}
\put(10554,-6961){\makebox(0,0)[lb]{\smash{\SetFigFont{8}{9.6}{bf}24}}}
\put(10201,-6961){\makebox(0,0)[lb]{\smash{\SetFigFont{8}{9.6}{bf}23}}}
\put(2926,-6961){\makebox(0,0)[lb]{\smash{\SetFigFont{8}{9.6}{bf}1}}}
\put(2626,-6961){\makebox(0,0)[lb]{\smash{\SetFigFont{8}{9.6}{bf}0}}}
\end{picture}

%% file: approxi.tex
\begin{picture}(0,0)%
\includegraphics{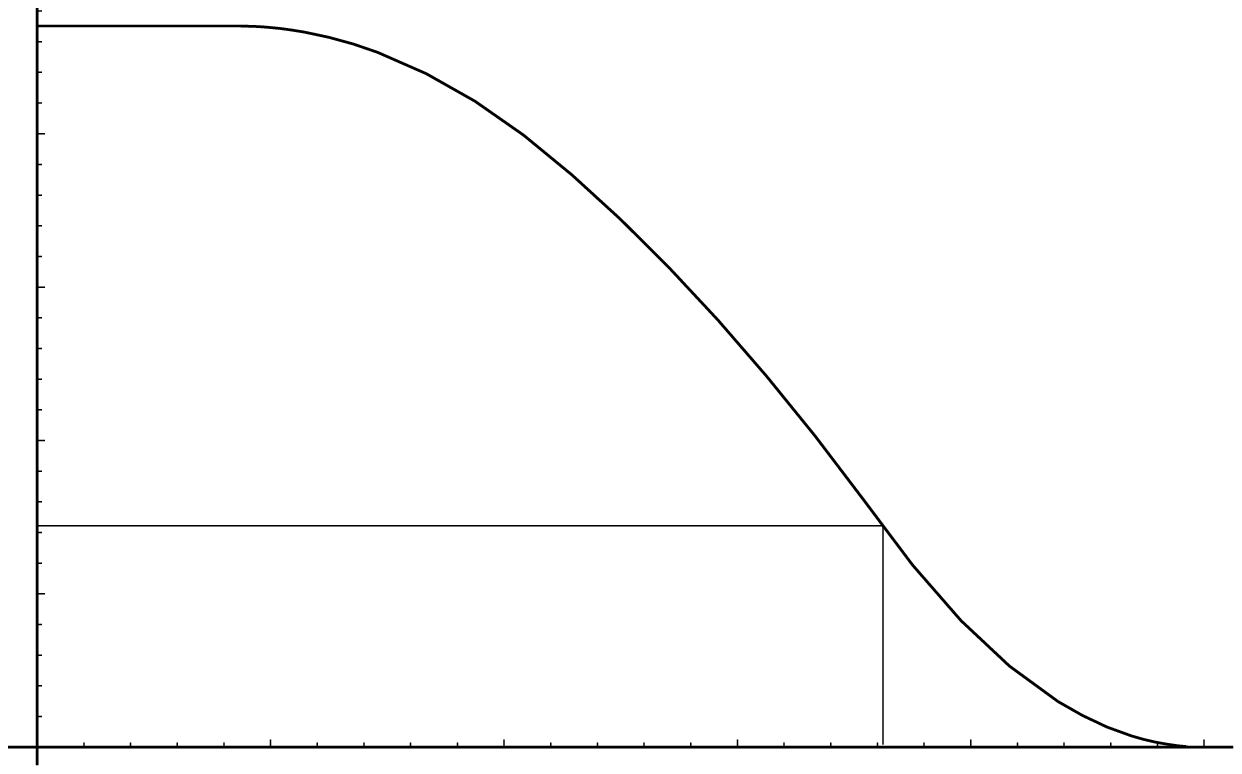}%
\end{picture}%
\setlength{\unitlength}{3158sp}%
\begingroup\makeatletter\ifx\SetFigFont\undefined
\def\x#1#2#3#4#5#6#7\relax{\def\x{#1#2#3#4#5#6}}%
\expandafter\x\fmtname xxxxxx\relax \def\y{splain}%
\ifx\x\y   
\gdef\SetFigFont#1#2#3{%
  \ifnum #1<17\tiny\else \ifnum #1<20\small\else
  \ifnum #1<24\normalsize\else \ifnum #1<29\large\else
  \ifnum #1<34\Large\else \ifnum #1<41\LARGE\else
     \huge\fi\fi\fi\fi\fi\fi
  \csname #3\endcsname}%
\else
\gdef\SetFigFont#1#2#3{\begingroup
  \count@#1\relax \ifnum 25<\count@\count@25\fi
  \def\x{\endgroup\@setsize\SetFigFont{#2pt}}%
  \expandafter\x
    \csname \romannumeral\the\count@ pt\expandafter\endcsname
    \csname @\romannumeral\the\count@ pt\endcsname
  \csname #3\endcsname}%
\fi
\fi\endgroup
\begin{picture}(7573,5289)(1184,-8644)
\put(8551,-8311){\makebox(0,0)[lb]{\smash{\SetFigFont{12}{14.4}{bf}1}}}
\put(1384,-4438){\makebox(0,0)[lb]{\smash{\SetFigFont{12}{14.4}{bf}2}}}
\put(1184,-5359){\makebox(0,0)[lb]{\smash{\SetFigFont{12}{14.4}{bf}1.5}}}
\put(1386,-6279){\makebox(0,0)[lb]{\smash{\SetFigFont{12}{14.4}{bf}1}}}
\put(1184,-7198){\makebox(0,0)[lb]{\smash{\SetFigFont{12}{14.4}{bf}0.5}}}
\put(1351,-3511){\makebox(0,0)[lb]{\smash{\SetFigFont{12}{14.4}{bf}$F(\eta)$}}}
\put(2776,-8311){\makebox(0,0)[lb]{\smash{\SetFigFont{12}{14.4}{bf}0.2}}}
\put(4201,-8311){\makebox(0,0)[lb]{\smash{\SetFigFont{12}{14.4}{bf}0.4}}}
\put(5626,-8311){\makebox(0,0)[lb]{\smash{\SetFigFont{12}{14.4}{bf}0.6}}}
\put(8326,-8611){\makebox(0,0)[lb]{\smash{\SetFigFont{12}{14.4}{bf}$\eta$}}}
\put(7051,-8311){\makebox(0,0)[lb]{\smash{\SetFigFont{12}{14.4}{bf}0.8}}}
\end{picture}

%% file: d3plot.tex
\begin{picture}(0,0)%
\special{psfile=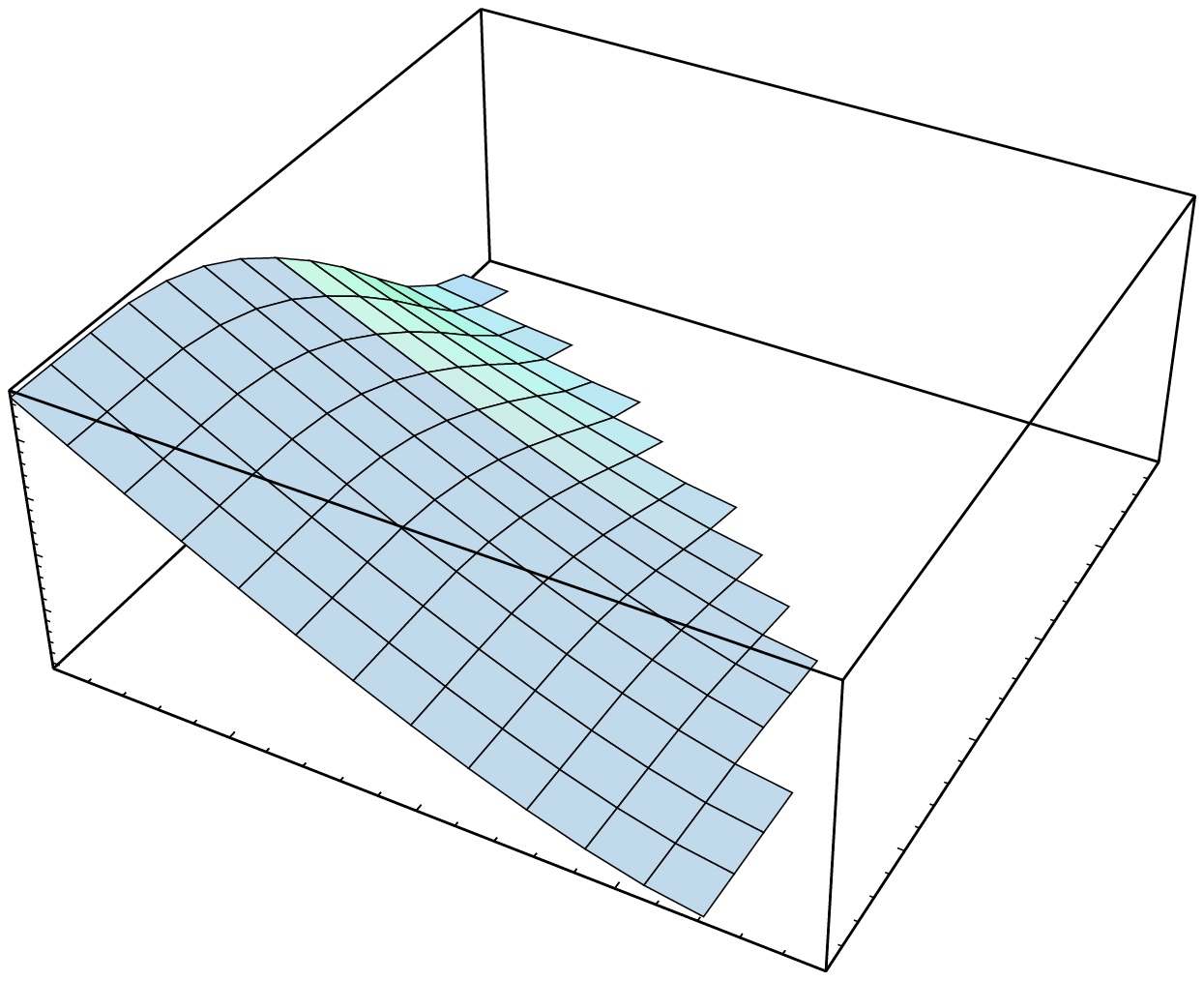}%
\end{picture}%
\setlength{\unitlength}{3158sp}%
\begingroup\makeatletter\ifx\SetFigFont\undefined
\def\x#1#2#3#4#5#6#7\relax{\def\x{#1#2#3#4#5#6}}%
\expandafter\x\fmtname xxxxxx\relax \def\y{splain}%
\ifx\x\y   
\gdef\SetFigFont#1#2#3{%
  \ifnum #1<17\tiny\else \ifnum #1<20\small\else
  \ifnum #1<24\normalsize\else \ifnum #1<29\large\else
  \ifnum #1<34\Large\else \ifnum #1<41\LARGE\else
     \huge\fi\fi\fi\fi\fi\fi
  \csname #3\endcsname}%
\else
\gdef\SetFigFont#1#2#3{\begingroup
  \count@#1\relax \ifnum 25<\count@\count@25\fi
  \def\x{\endgroup\@setsize\SetFigFont{#2pt}}%
  \expandafter\x
    \csname \romannumeral\the\count@ pt\expandafter\endcsname
    \csname @\romannumeral\the\count@ pt\endcsname
  \csname #3\endcsname}%
\fi
\fi\endgroup
\begin{picture}(8044,6333)(807,-9088)
\put(4857,-8885){\makebox(0,0)[lb]{\smash{\SetFigFont{12}{14.4}{bf}$v_{\perp}$}}}
\put(1324,-5446){\makebox(0,0)[lb]{\smash{\SetFigFont{12}{14.4}{bf}2}}}
\put(1182,-5807){\makebox(0,0)[lb]{\smash{\SetFigFont{12}{14.4}{bf}1.5}}}
\put(1440,-6156){\makebox(0,0)[lb]{\smash{\SetFigFont{12}{14.4}{bf}1}}}
\put(1292,-6495){\makebox(0,0)[lb]{\smash{\SetFigFont{12}{14.4}{bf}0.5}}}
\put(2476,-7561){\makebox(0,0)[lb]{\smash{\SetFigFont{12}{14.4}{bf}0.5}}}
\put(3826,-8011){\makebox(0,0)[lb]{\smash{\SetFigFont{12}{14.4}{bf}1}}}
\put(4951,-8536){\makebox(0,0)[lb]{\smash{\SetFigFont{12}{14.4}{bf}1.5}}}
\put(6376,-9061){\makebox(0,0)[lb]{\smash{\SetFigFont{12}{14.4}{bf}2}}}
\put(7201,-8086){\makebox(0,0)[lb]{\smash{\SetFigFont{12}{14.4}{bf}0.2}}}
\put(7651,-7411){\makebox(0,0)[lb]{\smash{\SetFigFont{12}{14.4}{bf}0.4}}}
\put(8401,-6211){\makebox(0,0)[lb]{\smash{\SetFigFont{12}{14.4}{bf}0.8}}}
\put(8776,-5686){\makebox(0,0)[lb]{\smash{\SetFigFont{12}{14.4}{bf}1}}}
\put(8026,-6811){\makebox(0,0)[lb]{\smash{\SetFigFont{12}{14.4}{bf}0.6}}}
\put(6676,-8836){\makebox(0,0)[lb]{\smash{\SetFigFont{12}{14.4}{bf}0}}}
\put(8851,-6586){\makebox(0,0)[lb]{\smash{\SetFigFont{12}{14.4}{bf}$\eta$}}}
\put(1576,-6961){\makebox(0,0)[lb]{\smash{\SetFigFont{12}{14.4}{bf}0}}}
\put(807,-4235){\makebox(0,0)[lb]{\smash{\SetFigFont{12}{14.4}{bf}$P(\eta,v_{\perp})$}}}
\end{picture}

%% file: auton1.tex
\begin{picture}(0,0)%
\special{psfile=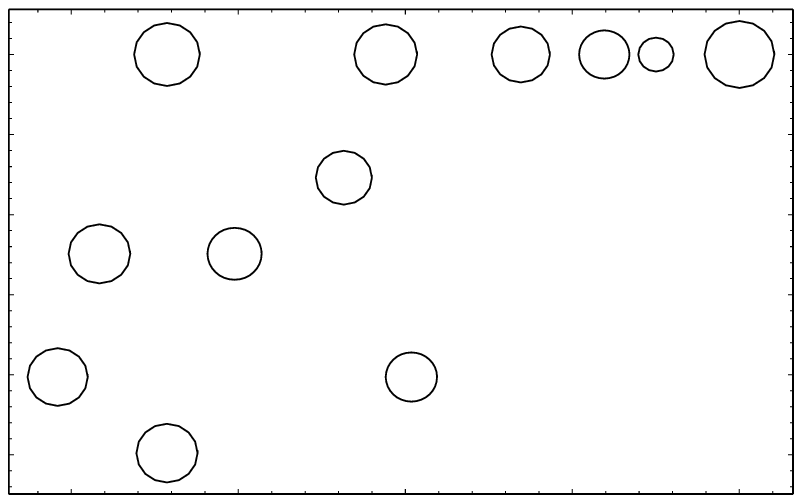}%
\end{picture}%
\setlength{\unitlength}{1973sp}%
\begingroup\makeatletter\ifx\SetFigFont\undefined
\def\x#1#2#3#4#5#6#7\relax{\def\x{#1#2#3#4#5#6}}%
\expandafter\x\fmtname xxxxxx\relax \def\y{splain}%
\ifx\x\y   
\gdef\SetFigFont#1#2#3{%
  \ifnum #1<17\tiny\else \ifnum #1<20\small\else
  \ifnum #1<24\normalsize\else \ifnum #1<29\large\else
  \ifnum #1<34\Large\else \ifnum #1<41\LARGE\else
     \huge\fi\fi\fi\fi\fi\fi
  \csname #3\endcsname}%
\else
\gdef\SetFigFont#1#2#3{\begingroup
  \count@#1\relax \ifnum 25<\count@\count@25\fi
  \def\x{\endgroup\@setsize\SetFigFont{#2pt}}%
  \expandafter\x
    \csname \romannumeral\the\count@ pt\expandafter\endcsname
    \csname @\romannumeral\the\count@ pt\endcsname
  \csname #3\endcsname}%
\fi
\fi\endgroup
\begin{picture}(8039,5072)(976,-8413)
\put(2701,-4561){\makebox(0,0)[lb]{\smash{\SetFigFont{12}{14.4}{bf}$VII$}}}
\put(4501,-5761){\makebox(0,0)[lb]{\smash{\SetFigFont{12}{14.4}{bf}$IV$}}}
\put(2101,-6361){\makebox(0,0)[lb]{\smash{\SetFigFont{12}{14.4}{bf}$VIII$}}}
\put(4876,-4561){\makebox(0,0)[lb]{\smash{\SetFigFont{12}{14.4}{bf}$III$}}}
\put(6226,-4561){\makebox(0,0)[lb]{\smash{\SetFigFont{12}{14.4}{bf}$II$}}}
\put(6976,-4561){\makebox(0,0)[lb]{\smash{\SetFigFont{12}{14.4}{bf}$I$}}}
\put(7501,-4561){\makebox(0,0)[lb]{\smash{\SetFigFont{12}{14.4}{bf}$I'$}}}
\put(8326,-4561){\makebox(0,0)[lb]{\smash{\SetFigFont{12}{14.4}{bf}$0$}}}
\put(3451,-6361){\makebox(0,0)[lb]{\smash{\SetFigFont{12}{14.4}{bf}$VI$}}}
\put(3451,-7861){\makebox(0,0)[lb]{\smash{\SetFigFont{12}{14.4}{bf}$IX$}}}
\put(5176,-7561){\makebox(0,0)[lb]{\smash{\SetFigFont{12}{14.4}{bf}$V$}}}
\put(1801,-7561){\makebox(0,0)[lb]{\smash{\SetFigFont{12}{14.4}{bf}$X$}}}
\put(976,-4636){\makebox(0,0)[lb]{\smash{\SetFigFont{12}{14.4}{bf}-1}}}
\put(976,-3811){\makebox(0,0)[lb]{\smash{\SetFigFont{12}{14.4}{bf}0}}}
\put(976,-5386){\makebox(0,0)[lb]{\smash{\SetFigFont{12}{14.4}{bf}-2}}}
\put(976,-6136){\makebox(0,0)[lb]{\smash{\SetFigFont{12}{14.4}{bf}-3}}}
\put(976,-6886){\makebox(0,0)[lb]{\smash{\SetFigFont{12}{14.4}{bf}-4}}}
\put(976,-7711){\makebox(0,0)[lb]{\smash{\SetFigFont{12}{14.4}{bf}-5}}}
\put(1951,-8386){\makebox(0,0)[lb]{\smash{\SetFigFont{12}{14.4}{bf}-8}}}
\put(3601,-8386){\makebox(0,0)[lb]{\smash{\SetFigFont{12}{14.4}{bf}-6}}}
\put(5176,-8386){\makebox(0,0)[lb]{\smash{\SetFigFont{12}{14.4}{bf}-4}}}
\put(6751,-8386){\makebox(0,0)[lb]{\smash{\SetFigFont{12}{14.4}{bf}-2}}}
\put(8476,-8386){\makebox(0,0)[lb]{\smash{\SetFigFont{12}{14.4}{bf}0}}}
\end{picture}

%% file: auton2.tex
\begin{picture}(0,0)%
\includegraphics{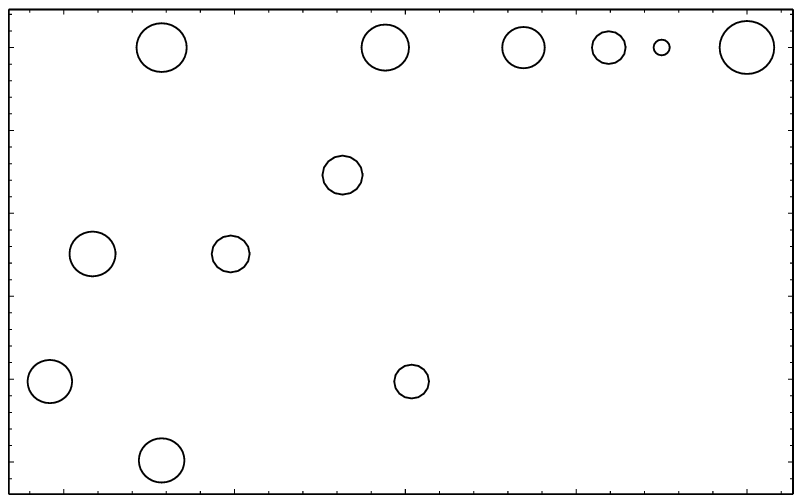}%
\end{picture}%
\setlength{\unitlength}{1973sp}%
\begingroup\makeatletter\ifx\SetFigFont\undefined
\def\x#1#2#3#4#5#6#7\relax{\def\x{#1#2#3#4#5#6}}%
\expandafter\x\fmtname xxxxxx\relax \def\y{splain}%
\ifx\x\y   
\gdef\SetFigFont#1#2#3{%
  \ifnum #1<17\tiny\else \ifnum #1<20\small\else
  \ifnum #1<24\normalsize\else \ifnum #1<29\large\else
  \ifnum #1<34\Large\else \ifnum #1<41\LARGE\else
     \huge\fi\fi\fi\fi\fi\fi
  \csname #3\endcsname}%
\else
\gdef\SetFigFont#1#2#3{\begingroup
  \count@#1\relax \ifnum 25<\count@\count@25\fi
  \def\x{\endgroup\@setsize\SetFigFont{#2pt}}%
  \expandafter\x
    \csname \romannumeral\the\count@ pt\expandafter\endcsname
    \csname @\romannumeral\the\count@ pt\endcsname
  \csname #3\endcsname}%
\fi
\fi\endgroup
\begin{picture}(8039,5147)(976,-8488)
\put(976,-3736){\makebox(0,0)[lb]{\smash{\SetFigFont{12}{14.4}{bf}0}}}
\put(976,-4561){\makebox(0,0)[lb]{\smash{\SetFigFont{12}{14.4}{bf}-1}}}
\put(976,-5386){\makebox(0,0)[lb]{\smash{\SetFigFont{12}{14.4}{bf}-2}}}
\put(976,-6961){\makebox(0,0)[lb]{\smash{\SetFigFont{12}{14.4}{bf}-4}}}
\put(976,-7786){\makebox(0,0)[lb]{\smash{\SetFigFont{12}{14.4}{bf}-5}}}
\put(1876,-8461){\makebox(0,0)[lb]{\smash{\SetFigFont{12}{14.4}{bf}-8}}}
\put(3526,-8461){\makebox(0,0)[lb]{\smash{\SetFigFont{12}{14.4}{bf}-6}}}
\put(5176,-8461){\makebox(0,0)[lb]{\smash{\SetFigFont{12}{14.4}{bf}-4}}}
\put(6826,-8461){\makebox(0,0)[lb]{\smash{\SetFigFont{12}{14.4}{bf}-2}}}
\put(8476,-8461){\makebox(0,0)[lb]{\smash{\SetFigFont{12}{14.4}{bf}0}}}
\put(976,-6136){\makebox(0,0)[lb]{\smash{\SetFigFont{12}{14.4}{bf}-3}}}
\end{picture}

%% file: auton3.tex
\begin{picture}(0,0)%
\includegraphics{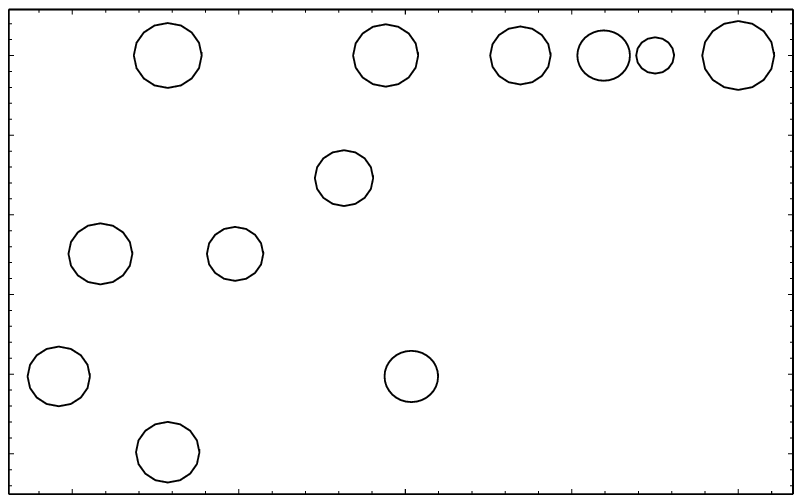}%
\end{picture}%
\setlength{\unitlength}{1973sp}%
\begingroup\makeatletter\ifx\SetFigFont\undefined
\def\x#1#2#3#4#5#6#7\relax{\def\x{#1#2#3#4#5#6}}%
\expandafter\x\fmtname xxxxxx\relax \def\y{splain}%
\ifx\x\y   
\gdef\SetFigFont#1#2#3{%
  \ifnum #1<17\tiny\else \ifnum #1<20\small\else
  \ifnum #1<24\normalsize\else \ifnum #1<29\large\else
  \ifnum #1<34\Large\else \ifnum #1<41\LARGE\else
     \huge\fi\fi\fi\fi\fi\fi
  \csname #3\endcsname}%
\else
\gdef\SetFigFont#1#2#3{\begingroup
  \count@#1\relax \ifnum 25<\count@\count@25\fi
  \def\x{\endgroup\@setsize\SetFigFont{#2pt}}%
  \expandafter\x
    \csname \romannumeral\the\count@ pt\expandafter\endcsname
    \csname @\romannumeral\the\count@ pt\endcsname
  \csname #3\endcsname}%
\fi
\fi\endgroup
\begin{picture}(8039,5147)(976,-8488)
\put(976,-3811){\makebox(0,0)[lb]{\smash{\SetFigFont{12}{14.4}{bf}0}}}
\put(976,-4636){\makebox(0,0)[lb]{\smash{\SetFigFont{12}{14.4}{bf}-1}}}
\put(976,-5386){\makebox(0,0)[lb]{\smash{\SetFigFont{12}{14.4}{bf}-2}}}
\put(976,-6136){\makebox(0,0)[lb]{\smash{\SetFigFont{12}{14.4}{bf}-3}}}
\put(976,-6886){\makebox(0,0)[lb]{\smash{\SetFigFont{12}{14.4}{bf}-4}}}
\put(976,-7711){\makebox(0,0)[lb]{\smash{\SetFigFont{12}{14.4}{bf}-5}}}
\put(1951,-8461){\makebox(0,0)[lb]{\smash{\SetFigFont{12}{14.4}{bf}-8}}}
\put(3601,-8461){\makebox(0,0)[lb]{\smash{\SetFigFont{12}{14.4}{bf}-6}}}
\put(5176,-8461){\makebox(0,0)[lb]{\smash{\SetFigFont{12}{14.4}{bf}-4}}}
\put(6751,-8461){\makebox(0,0)[lb]{\smash{\SetFigFont{12}{14.4}{bf}-2}}}
\put(8401,-8461){\makebox(0,0)[lb]{\smash{\SetFigFont{12}{14.4}{bf}0}}}
\end{picture}

%% file: berggflaabs.tex
\begin{picture}(0,0)%
\includegraphics{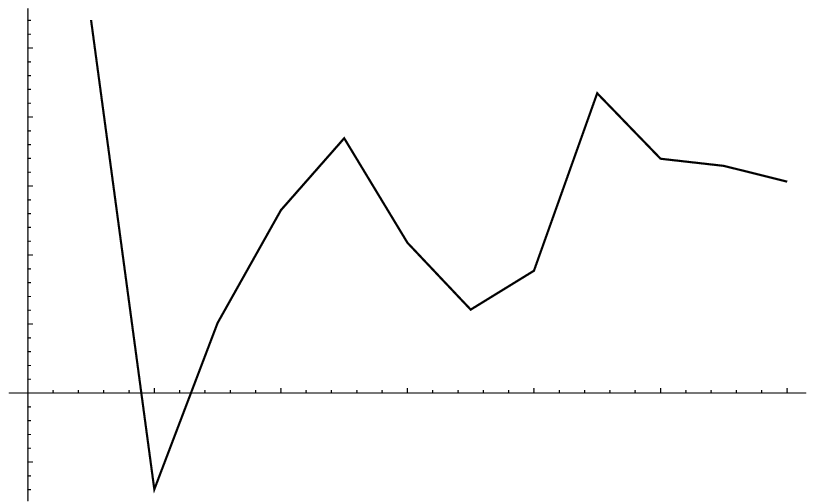}%
\end{picture}%
\setlength{\unitlength}{2368sp}%
\begingroup\makeatletter\ifx\SetFigFont\undefined
\def\x#1#2#3#4#5#6#7\relax{\def\x{#1#2#3#4#5#6}}%
\expandafter\x\fmtname xxxxxx\relax \def\y{splain}%
\ifx\x\y   
\gdef\SetFigFont#1#2#3{%
  \ifnum #1<17\tiny\else \ifnum #1<20\small\else
  \ifnum #1<24\normalsize\else \ifnum #1<29\large\else
  \ifnum #1<34\Large\else \ifnum #1<41\LARGE\else
     \huge\fi\fi\fi\fi\fi\fi
  \csname #3\endcsname}%
\else
\gdef\SetFigFont#1#2#3{\begingroup
  \count@#1\relax \ifnum 25<\count@\count@25\fi
  \def\x{\endgroup\@setsize\SetFigFont{#2pt}}%
  \expandafter\x
    \csname \romannumeral\the\count@ pt\expandafter\endcsname
    \csname @\romannumeral\the\count@ pt\endcsname
  \csname #3\endcsname}%
\fi
\fi\endgroup
\begin{picture}(7159,3966)(1201,-7908)
\put(1201,-4336){\makebox(0,0)[lb]{\smash{\SetFigFont{12}{14.4}{bf}2.25}}}
\put(1201,-4861){\makebox(0,0)[lb]{\smash{\SetFigFont{12}{14.4}{bf}2}}}
\put(1201,-5386){\makebox(0,0)[lb]{\smash{\SetFigFont{12}{14.4}{bf}1.75}}}
\put(1201,-5986){\makebox(0,0)[lb]{\smash{\SetFigFont{12}{14.4}{bf}1.5}}}
\put(1201,-6511){\makebox(0,0)[lb]{\smash{\SetFigFont{12}{14.4}{bf}1.25}}}
\put(1201,-7636){\makebox(0,0)[lb]{\smash{\SetFigFont{12}{14.4}{bf}0.75}}}
\put(3076,-7561){\makebox(0,0)[lb]{\smash{\SetFigFont{12}{14.4}{bf}I'}}}
\put(4126,-7561){\makebox(0,0)[lb]{\smash{\SetFigFont{12}{14.4}{bf}II}}}
\put(5101,-7561){\makebox(0,0)[lb]{\smash{\SetFigFont{12}{14.4}{bf}IV}}}
\put(6076,-7561){\makebox(0,0)[lb]{\smash{\SetFigFont{12}{14.4}{bf}VI}}}
\put(7126,-7561){\makebox(0,0)[lb]{\smash{\SetFigFont{12}{14.4}{bf}VIII}}}
\put(8101,-7561){\makebox(0,0)[lb]{\smash{\SetFigFont{12}{14.4}{bf}X}}}
\end{picture}

%% file: bergkflaabs.tex
\begin{picture}(0,0)%
\includegraphics{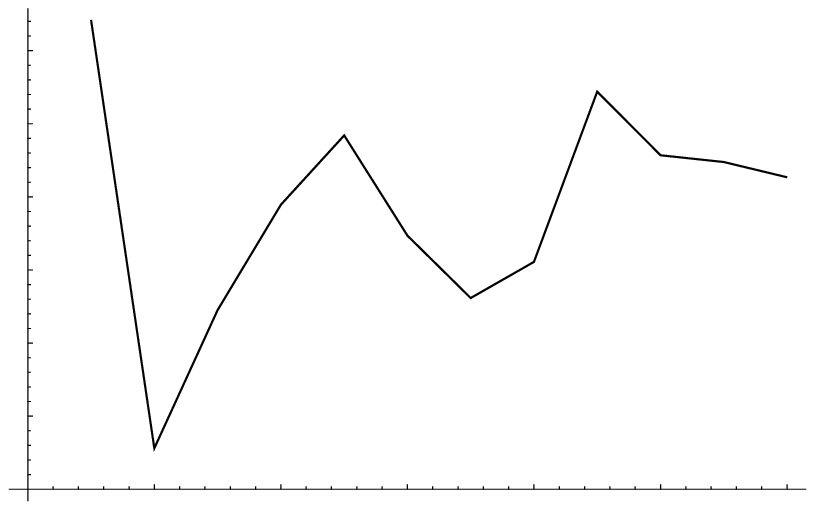}%
\end{picture}%
\setlength{\unitlength}{2368sp}%
\begingroup\makeatletter\ifx\SetFigFont\undefined
\def\x#1#2#3#4#5#6#7\relax{\def\x{#1#2#3#4#5#6}}%
\expandafter\x\fmtname xxxxxx\relax \def\y{splain}%
\ifx\x\y   
\gdef\SetFigFont#1#2#3{%
  \ifnum #1<17\tiny\else \ifnum #1<20\small\else
  \ifnum #1<24\normalsize\else \ifnum #1<29\large\else
  \ifnum #1<34\Large\else \ifnum #1<41\LARGE\else
     \huge\fi\fi\fi\fi\fi\fi
  \csname #3\endcsname}%
\else
\gdef\SetFigFont#1#2#3{\begingroup
  \count@#1\relax \ifnum 25<\count@\count@25\fi
  \def\x{\endgroup\@setsize\SetFigFont{#2pt}}%
  \expandafter\x
    \csname \romannumeral\the\count@ pt\expandafter\endcsname
    \csname @\romannumeral\the\count@ pt\endcsname
  \csname #3\endcsname}%
\fi
\fi\endgroup
\begin{picture}(7159,4272)(1201,-8188)
\put(1201,-4336){\makebox(0,0)[lb]{\smash{\SetFigFont{12}{14.4}{bf}1.2}}}
\put(1201,-4861){\makebox(0,0)[lb]{\smash{\SetFigFont{12}{14.4}{bf}1}}}
\put(1201,-5461){\makebox(0,0)[lb]{\smash{\SetFigFont{12}{14.4}{bf}0.8}}}
\put(1201,-6061){\makebox(0,0)[lb]{\smash{\SetFigFont{12}{14.4}{bf}0.6}}}
\put(1201,-6661){\makebox(0,0)[lb]{\smash{\SetFigFont{12}{14.4}{bf}0.4}}}
\put(1201,-7261){\makebox(0,0)[lb]{\smash{\SetFigFont{12}{14.4}{bf}0.2}}}
\put(3076,-8161){\makebox(0,0)[lb]{\smash{\SetFigFont{12}{14.4}{bf}I'}}}
\put(4051,-8161){\makebox(0,0)[lb]{\smash{\SetFigFont{12}{14.4}{bf}II}}}
\put(5101,-8161){\makebox(0,0)[lb]{\smash{\SetFigFont{12}{14.4}{bf}IV}}}
\put(6151,-8161){\makebox(0,0)[lb]{\smash{\SetFigFont{12}{14.4}{bf}VI}}}
\put(7126,-8161){\makebox(0,0)[lb]{\smash{\SetFigFont{12}{14.4}{bf}VIII}}}
\put(8101,-8161){\makebox(0,0)[lb]{\smash{\SetFigFont{12}{14.4}{bf}X}}}
\end{picture}

%% file: gittflaabs.tex
\begin{picture}(0,0)%
\includegraphics{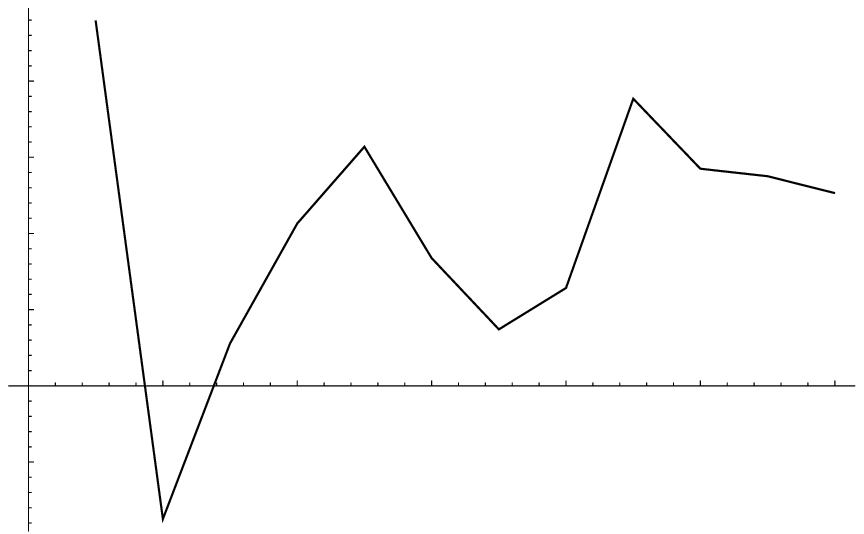}%
\end{picture}%
\setlength{\unitlength}{2368sp}%
\begingroup\makeatletter\ifx\SetFigFont\undefined
\def\x#1#2#3#4#5#6#7\relax{\def\x{#1#2#3#4#5#6}}%
\expandafter\x\fmtname xxxxxx\relax \def\y{splain}%
\ifx\x\y   
\gdef\SetFigFont#1#2#3{%
  \ifnum #1<17\tiny\else \ifnum #1<20\small\else
  \ifnum #1<24\normalsize\else \ifnum #1<29\large\else
  \ifnum #1<34\Large\else \ifnum #1<41\LARGE\else
     \huge\fi\fi\fi\fi\fi\fi
  \csname #3\endcsname}%
\else
\gdef\SetFigFont#1#2#3{\begingroup
  \count@#1\relax \ifnum 25<\count@\count@25\fi
  \def\x{\endgroup\@setsize\SetFigFont{#2pt}}%
  \expandafter\x
    \csname \romannumeral\the\count@ pt\expandafter\endcsname
    \csname @\romannumeral\the\count@ pt\endcsname
  \csname #3\endcsname}%
\fi
\fi\endgroup
\begin{picture}(7449,4210)(901,-8014)
\put(901,-4411){\makebox(0,0)[lb]{\smash{\SetFigFont{12}{14.4}{bf}2}}}
\put(901,-5086){\makebox(0,0)[lb]{\smash{\SetFigFont{12}{14.4}{bf}1.75}}}
\put(901,-5686){\makebox(0,0)[lb]{\smash{\SetFigFont{12}{14.4}{bf}1.5}}}
\put(901,-6286){\makebox(0,0)[lb]{\smash{\SetFigFont{12}{14.4}{bf}1.25}}}
\put(901,-7486){\makebox(0,0)[lb]{\smash{\SetFigFont{12}{14.4}{bf}0.75}}}
\put(2776,-7261){\makebox(0,0)[lb]{\smash{\SetFigFont{12}{14.4}{bf}I'}}}
\put(3826,-7261){\makebox(0,0)[lb]{\smash{\SetFigFont{12}{14.4}{bf}II}}}
\put(4876,-7261){\makebox(0,0)[lb]{\smash{\SetFigFont{12}{14.4}{bf}IV}}}
\put(6001,-7261){\makebox(0,0)[lb]{\smash{\SetFigFont{12}{14.4}{bf}VI}}}
\put(6976,-7261){\makebox(0,0)[lb]{\smash{\SetFigFont{12}{14.4}{bf}VIII}}}
\put(8101,-7261){\makebox(0,0)[lb]{\smash{\SetFigFont{12}{14.4}{bf}X}}}
\end{picture}

%% file: surf3.bbl
\begin{thebibliography}{99}

\bibitem{BAA}  Baake M,  Kramer P, Schlottmann M and  Zeidler D,
                 {\em Int. J. Mod. Phys.} {\bf B4} (1990) 2217-67 
\bibitem{BO}   de Boissieu M, Stephens P, Boudard M, Janot C, Chapman D L
and Audier M, 
{\em J. Phys. Condens. Matter} {\bf 6} (1994) 10725
\bibitem{EB} Ebert Ph, Yue F, and Urban K, {\em Phys. Rev.} {\bf B 57}
(1998) 2821
\bibitem{elser}  Elser V, {\em Phil.\ Mag.\ } B{\bf 73} (1996) 641
\bibitem{GI1} Gierer M, Van Hove M A, Goldman A I, Shen Z, Chang S-L, 
Jenks C-M, Zhang C-M, Thiel P A, {\em Phys. Rev. Lett.}{\bf 78} (1997) 467
\bibitem{GI2} Gierer M, Van Hove M A, Goldman A I, Shen Z, Chang S-L, 
Pinhero P J, Jenks C M, Anderegg J W, Zhang C-M, Thiel P A, 
{\em Phys. Rev.}{\bf B 57} (1998) 7628
\bibitem{KA} Katz A and Gratias D, in: {\em Proc. 5th Int. Conf.
on Quasicrystals}, eds. C Janot and R Mosseri, World Scientific, 
Singapore 1995, pp. 164-167
                 
\bibitem{KE} Kepler J (1619) {\em Harmonice Mundi}, Germ. Transl.
{\em Weltharmonik} by M Caspar, p. 74, Oldenbourg, Munich 1973

\bibitem{KR0} Kramer P and Neri R, 
                 {\em Acta Cryst.} ${\bf A\; 40}$ (1984), 580-587



\bibitem{KR} Kramer P, Papadopolos Z, Zeidler D, 
{\em Symmetries of Icosahedral Quasicrystals}, in: 
             {\em Symmetry in Science V}, eds B. Gruber,
L. C. Biedenharn and H. D. Doebner, Plenum, New York (1991), 395-427
              
\bibitem{KR1} Kramer P, Papadopolos Z and  
              Liebermeister W,
              in: {\em Proc. 6th Int. Conf. on Quasicrystals}, eds.
              S. Takeuchi and T. Fujiwara, World Scientific, Singapore
              (1998), 71-76\\ 
              Kramer P, Papadopolos Z and Liebermeister W,
                 {\em Atomic Positions for the Icosahedral F--Phase Tiling} 
                 in: Proc. Aperiodic '97, eds. 

\bibitem{KR2} Kasner G, Papadopolos Z, Kramer P and B\"urgler D E,
              submitted for publication

\bibitem{KR3} Papadopolos Z, Kramer P and Zeidler D, 
              {\em J. Non-Cryst. Solids} {\bf 153,154} (1993), 215-220 

\bibitem{MO}  Moody R V, {\em Meyer Sets and Their Duals}, 
              in: {\em The Mathematics of Long-Range Aperiodic
              Order}, Ed. R. V. Moody, Kluwer, Dordrecht 1997, pp. 403-441


\bibitem{BU} Schaub T M, B\"urgler D E, 
                 G\"untherodt H-J, and Suck J B,  
                 {\em Phys.Rev.Lett.} {\bf 73} (1994) 1255-8,
                 \\ 
                 Schaub T M , B\"urgler D E, 
                 G\"untherodt H-J, Suck J B and Audier M
                 {\em Appl. Phys. A} {\bf 61} (1995) 491-501 
\end{thebibliography}
